\documentclass[
showkeys,reprint,
nofootinbib,
 amsmath,amssymb,
 aps,
floatfix
]{revtex4-1}
\usepackage[varg]{txfonts}
\usepackage{graphicx}
\usepackage{natbib}
\usepackage{bm}%
\usepackage{xcolor,colortbl}
\usepackage{booktabs}
\usepackage{wrapfig}
\usepackage{setspace}
\usepackage{cellspace}
\usepackage{latexsym,color}
\usepackage{amssymb}
\usepackage[T1]{fontenc}
\usepackage{amsmath,amssymb,mathrsfs,amsfonts,dsfont}
\usepackage{color}
\usepackage{xcolor}
\definecolor{LinkBlue}{rgb}{0.00,0.00,1.00}
\definecolor{Grree}{rgb}{0.22,0.71,0.06}
\definecolor{Oran}{rgb}{1.00,0.50,0.25}

\usepackage[colorlinks=true,linkcolor=LinkBlue,urlcolor=Grree,	 citecolor=LinkBlue,citecolor=LinkBlue,hyperfootnotes=true]{hyperref}

\usepackage{enumitem}

\definecolor{lightgray}{rgb}{0.75,0.75,0.75}

\def\be{\begin{equation}}
\def\ee{\end{equation}}
\def\bea{\begin{eqnarray}}

\newcommand{\apss}{Ap\&SS}
\newcommand{\apjs}{APJS}
\newcommand{\mnras}{MNRAS}

\def\eea{\end{eqnarray}}

\newcommand{\il}{~}

\count\footins = 1000

\begin{document}

\title{Proto-jets configurations  in   \textbf{RADs}   orbiting a   Kerr \textbf{SMBH}: symmetries and   limiting surfaces }
\author{D. Pugliese\& Z. Stuchl\'{\i}k}

\affiliation{
Institute of Physics and Research Centre of Theoretical Physics and Astrophysics, Faculty of Philosophy\&Science,
 Bezru\v{c}ovo n\'{a}m\v{e}st\'{i} 13, CZ-74601 Opava, Czech Republic\\
 \email{d.pugliese.physics@gmail.com;zdenek.stuchlik@physics.cz} }

 \date{Received XXXX xx,XXXX; accepted YYYY yy, YYYY}

% \abstract{}{}{}{}{}
% 5 {} token are mandatory

  \begin{abstract}
Ringed accretion disks (\textbf{RADs}) are agglomerations of perfect-fluid tori orbiting around a single central attractor that could arise during complex matter inflows in active galactic nuclei. We focus our analysis to axi-symmetric accretion tori orbiting in the equatorial plane of a supermassive Kerr black hole; equilibrium configurations, possible instabilities, and evolutionary sequences of \textbf{RADs} were discussed in our previous works. In the present work we discuss special instabilities related to open equipotential surfaces governing the material funnels emerging at various regions of the \textbf{RADs}, being located between two or more individual toroidal configurations of the agglomerate. These open structures could be associated to proto-jets. Boundary limiting surfaces are highlighted, connecting the emergency of the jet-like instabilities with the black hole dimensionless spin. These instabilities are observationally significant for active galactic nuclei, being related to outflows of matter in jets emerging from more than one torus of \textbf{RADs} orbiting around supermassive black holes.
\end{abstract}
\keywords{
Black hole physics -- jets--Gravitation -- Hydrodynamics -- Accretion, accretion disks -- Galaxies: active -- Galaxies: jets}
\date{\today}%Accretion disks, accretion, jets, black hole physics, hydrodynamics

\maketitle
%Uncomment for Submitted to journal title message
%\maketitle

%
\maketitle

\section{Introduction}
Spacetime symmetries play a dominant role in  Astrophysics. The case of accretion disks,  and especially the  constrained    axi-symmetric (coplanar) tori orbiting a central Kerr supermassive black hole (\textbf{SMBH}) represents an example   where both stable and unstable configurations can be predominantly determined by the geometric properties of the background.

More generally the physics of accretion disks is regulated by the balance of many factors influencing their  evolution and morphology; magnetic fields, thermal or viscous  processes  for example play a combined and different role,  from  the phase of disk formation to accretion,  eventually characterizing   different accretion disk models.
Nevertheless it is possible to trace a correlation between  accretion disk models, determined by this special balance and the disk attractor itself.
In the  Polish-doughnut (P-D) thick accretion disk  model  \cite{AbraFra},
 the role played by the curvature effects and spacetime symmetries is essential  in setting constraints  on   axi-symmetric   accretion tori and in these situations,
the geometric properties of the spacetime  become relevant with respect to other ingredients in the determination of the disk forces balance.
In here we consider ringed accretion disks (\textbf{RADs}), introduced  in \cite{pugtot} and detailed in \cite{ringed,open,dsystem,Long}, featuring the formation and interaction  of several, both corotating and counterrotating, axi-symmetric (coplanar) tori orbiting a central Kerr \textbf{SMBH}.
%\btb{In particular  while equiliobr\cite{ringed}}

These structures  may form around  \textbf{SMBH} in active galactic nucleai (\textbf{AGNs}), where the attractor, interacting  with its environment  during its life-time can give rise  to different accretion periods.
The case of a single orbiting  torus  stands as   limiting case  of the \textbf{RAD}.

The spacetime symmetries  determined by the central attractor strongly constrain  the  configurations existence and stability.
 Properties of each  torus are  determined by an effective potential function
enclosing  the background Kerr geometry and the centrifugal effects. Equipotential surfaces,  being   also equipressure surfaces, are  associated with critical points identifying the toroidal surfaces of the disk.
The cusped surfaces are the critical topologies associated to the torus  unstable  phases.
The outflow of matter through the cusp occurs  by
the Paczynski  mechanism of violation of mechanical equilibrium of the tori \cite{PW80}, i.e.  an
instability in the balance of the gravitational and inertial forces and the pressure gradients in the fluid \cite{AbraFra}.
The instabilities  inside the ringed accretion disk may give rise to accretion into the attractor,  whereas the open equipotential surfaces have  been related   to  ``proto jet-shell'' structures, a  general discussion   of these special config1urations can be found for example in \cite{KJA78,AJS78,Sadowski:2015jaa,Lasota:2015bii,Lyutikov(2009),Blaschke:2016uyo,2017PhRvD..96j4050S,Madau(1988),Sikora(1981),Desitte,Salny}. The   interaction between  several   launching points of  jets, or  other topologies of the decompositions,  leads to collision in a  couple, inducing a second instability phase with    accretion  or a further  proto-jet formation, finally  a phase,  or  a cycle,   of a ``drying-feeding'' process   between two \textbf{RADs} sub-configurations.
In this work  we concentrate our  attention on these   open configurations and their constraints.

From phenomenological viewpoint, the dynamics of the unstable phases
of this system is significant for the high energy phenomena related to accretion onto super-massive
black holes in \textbf{AGNs}, and the extremely energetic phenomena in quasars which
could be observable in their X-ray emission, as  the X-ray obscuration and absorption by one of the \textbf{RAD} ring.
The radially oscillating tori of the ringed disk could be related to the high-frequency quasi periodic oscillations
observed in non-thermal X-ray emission from compact objects (QPOs), a still obscure feature of the X-ray astronomy
related to the inner parts of the disk.
\textbf{BH} accretion rings models may be  revealed by future X-ray spectroscopy, from the
 study of  relatively indistinct excesses
on top of the relativistically broadened spectral line  profile \cite{S11etal,KS10,Schee:2008fc}. The predicted relatively indistinct excesses
 of the relativistically broadened  emission-line components, shall  arise in a well-confined
radial distance in the accretion disk, envisaging  therefore a sort of rings model which may be adapted as a special case  of the  model discussed in \cite{ringed,open}.

In Section(\ref{Sec:models}) we introduce  the \textbf{RADs} model  discussing    the tori  morphology, Sec.\il(\ref{Sec:RADs}) and Sec.\il(\ref{Sec:RADmorpho}), while  instabilities are  considered   in   Section\il(\ref{Sec:insta-Rad-Proc}).
{Geometrical correlations of unstable configurations} are discussed in Sec.\il(\ref{Sec:geometrical-corre}). %
Section\il(\ref{Sec:div-spee})  is devoted to the
parameter setting and  discussion on the system rotational symmetries, in this section we provide also   exact form of  fluid specific  angular momentum $\breve{\ell}$ introduced in  \cite{open} and also considered in \cite{Long}.
{Analysis of the   open configurations} is in Section(\ref{Sec:lambda-right}), while  in  Section (\ref{Sec:det-open-in-ext})
restrictions on the existence of the open configurations are investigated introducing some  {limiting} surfaces.
Concluding remarks follow in  Section\il(\ref{Sec:Conclusion}).
\section{Tori in the Kerr spacetime}\label{Sec:models}
We start by presenting the Kerr  metric tensor  in the Boyer-Lindquist (BL)  coordinates\footnote{We adopt the
geometrical  units $c=1=G$ and  the $(-,+,+,+)$ signature, Greek indices run in $\{0,1,2,3\}$.  The   four-velocity  satisfy $u^{\alpha} u_{\alpha}=-1$. The radius $r$ has unit of
mass $[M]$, and the angular momentum  units of $[M]^2$, the velocities  $[u^t]=[u^r]=1$
and $[u^{\varphi}]=[u^{\theta}]=[M]^{-1}$ with $[u^{\varphi}/u^{t}]=[M]^{-1}$ and
$[u_{\varphi}/u_{t}]=[M]$. For the seek of convenience, we always consider the
dimensionless  energy and effective potential $[V_{eff}]=1$ and an angular momentum per
unit of mass $[L]/[M]=[M]$.}.
\( \{t,r,\theta ,\phi \}\)
\bea \nonumber&& ds^2=-dt^2+\frac{\rho^2}{\Delta}dr^2+\rho^2
d\theta^2+(r^2+a^2)\sin^2\theta
d\phi^2+
\\\label{alai}
&&\frac{2M}{\rho^2}r(dt-a\sin^2\theta d\phi)^2\ ,
\\
\nonumber
&&
 \mbox{where}\quad\rho^2\equiv r^2+a^2\cos\theta^2\quad \mbox{and } \quad \Delta\equiv r^2-2 M r+a^2,
\eea
$a=J/M\in[0,1]$ is  the specific angular momentum,  $M$  a mass parameter and  $J$ is the
total angular momentum of the gravitational source. The non-rotating  limiting case $a=0$ is the   Schwarzschild metric while the extreme Kerr black hole  has dimensionless spin $a/M=1$.
Radii
{
\bea&&
r_{h}\equiv M+\sqrt{M^2-a^2};\quad r_{c}\equiv M-\sqrt{M^2-a^2}\\
  &&r_{\epsilon}^{+}\equiv M+\sqrt{M^2- a^2 \cos\theta^2};
%&&\nonumber
\eea
are the  horizons $r_c<r_h$ and the outer static limit $r_{\epsilon}^+$  respectively,
being  $r_h<r_{\epsilon}^+$ on   $\theta\neq0$  and  $r_{\epsilon}^+=2M$  in the equatorial plane $\theta=\pi/2$.}

   Ringed accretion disks are toroidal  configurations of  perfect fluids orbiting a   central  Kerr black hole \textbf{(BH)} attractor.
Quantities
\be\label{Eq:after}
{E} \equiv -g_{\alpha \beta}\xi_{t}^{\alpha} p^{\beta},\quad L \equiv
g_{\alpha \beta}\xi_{\phi}^{\alpha}p^{\beta}\ ,
\ee
are  constants of motion, where the covariant
components $p_{\phi}$ and $p_{t}$ of a particle four--momentum are
conserved along the   geodesics.
The constant $L$ in Eq.\il(\ref{Eq:after}) may be interpreted       as the axial component of the angular momentum  of a   particle for
timelike geodesics and $E$ as representing the total energy of the test particle
 coming from radial infinity, as measured  by  a static observer at infinity.
This is a property derived from the presence of the two Killing vector fields     $\xi_{\phi}=\partial_{\phi} $, rotational Killing field,  and  $\xi_{t}=\partial_{t} $,  which is
the Killing field representing the stationarity of the  spacetime.
 In  the test particle circular motion one  can  limit the  analysis to the case of  positive values of $a$
for corotating  $(L>0)$ and counterrotating   $(L<0)$ orbits with respect to the black hole. In fact the
 metric tensor (\ref{alai}) is  invariant under the application of any two different transformations: $x^\alpha\rightarrow-x^\alpha$
 {where $x^{\alpha}=(t,\phi)$, thus the metric is invariant for exchange
   of couple $(t, \phi)\rightarrow(-t, -\phi)$, or   $(a,t)\rightarrow(-a, -t)$, or after a change
   $(a, \phi)\rightarrow(-a, -\phi)$}, consequently,    the    test particle dynamics is invariant under the mutual transformation of the parameters
$(a,L)\rightarrow(-a,-L)$.

It will be important to consider in the analysis of the ringed disks  the   notable  radii  $r_{\mathcal{N}}^{\pm}\equiv \{r_{\gamma}^{\pm}, r_{mbo}^{\pm},r_{mso}^{\pm}\}$,  defining   the geodesic structure of the Kerr spacetime  with respect to  the matter distribution.
Specifically, for timelike particle orbits, $r_{\gamma}^{\pm}$ is  the \emph{marginally circular orbit}  or  the photon circular orbit, timelike  circular orbits  can fill  the spacetime region $r>r_{\gamma}^{\pm}$. The \emph{marginally stable circular orbit}  $r_{mso}^{\pm}$: stable orbits are in $r>r_{mso}^{\pm}$ for counterrotating and corotating particles respectively.  The \emph{marginally  bounded circular  orbit}  is $r_{mbo}^{\pm}$, where
 $E_{\pm}(r_{mbo}^{\pm})=1$  \cite{Pu:Kerr,Pu:KN,Pu:class,Pu:Charged,KNcharged,Stuchlik:2004wk}.
 {The  effective potential, regulating the motion of  test particle circular geodesics can admit, as function of $r/M$, minimum points  correspondent to  particle stable circular  orbits only in the region $r>r_{mso}^{\pm}$,  respectively for counterrotating  and corotating motion with respect to the central Kerr black hole; the effective potential admits the maximum points, correspondent to  particle unstable circular  orbits, only in the region $r\in]r_{\gamma}^{\pm},r_{mso}^{\pm}[$.
 The energy $E$ of the particle, as measured by an observer at infinity, will be greater that
 the value at infinity $E=\mu$, for the  particle of mass $\mu$   on orbits in  $r\in]r^{\pm}_{\gamma}, r^{\pm}_{mbo}[$,
 and the energy will be lower then limiting  $E=\mu$, in  the region $r\in]r^{\pm}_{mbo}, r^{\pm}_{mso}[$.}

The geodesic structure represents a geometric property of the spacetime    consisting of  the   union of the   orbital regions with boundaries in $r_{\mathcal{N}}$  \cite{ergon,open,dsystem}.
It can be decomposed, for $a\neq0$, into  the geodesic structures for corotating ($r_{\mathcal{N}}^-$) and counterrotating  ($r_{\mathcal{N}}^+$) matter {according to the convention adopted here and in \cite{ringed}}.
Given $r_i\in r_{\mathcal{N}}^{\pm}$,  we adopt  the  notation for any function $\mathbf{Q}(r):\;\mathbf{Q}_i\equiv\mathbf{Q}(r_i)$, therefore for example $\ell_{mso}^+\equiv\ell_+(r_{mso}^+)$, and more generally given the radius  $r_{\ast}$ and the function  $\mathbf{Q}(r)$,  there is $\mathbf{Q}_{\ast}\equiv\mathbf{Q}(r_{\ast})$.

For the
symmetries of the problem, we  assume $\partial_t \mathbf{Q}=0$ and
$\partial_{\varphi} \mathbf{Q}=0$,  with $\mathbf{Q}$ being a generic spacetime tensor \cite{Pugliese:2011aa,pugtot}, and
a   one-species particle perfect  fluid system    described by  the  energy momentum tensor
\be\label{E:Tm}
T_{\alpha \beta}=(\varrho +p) u_{\alpha} u_{\beta}+\  p g_{\alpha \beta},
\ee
where $\varrho$ and $p$ are  the total energy density and
pressure, respectively, as measured by an observer comoving with the fluid whose four-velocity $u^{\alpha}$  is
a timelike flow vector field.
\emph{Continuity  equation} and the \emph{Euler equation} are  respectively:
\bea\label{E:1a0}
u^\alpha\nabla_\alpha\varrho+(p+\varrho)\nabla^\alpha u_\alpha=0\,\\
\nonumber
%\label{Eulerif0}
(p+\varrho)u^\alpha \nabla_\alpha u^\gamma+ \ h^{\beta\gamma}\nabla_\beta p=0\,\
\\\nonumber h_{\alpha \beta}=g_{\alpha \beta}+ u_\alpha u_\beta,\quad (\nabla_\alpha g_{\beta\gamma}=0)
\eea

We consider   the  fluid toroidal configurations (with  $u^{\theta}=0$) centered on  the  plane $\theta=\pi/2$, and  defined by the constraint
$u^r=0$.
Assuming a barotropic equation of state $p=p(\varrho)$,
the Euler  equation (\ref{E:1a0}) provides the following equation
%\btb
{
%f
\bea&&\label{Eq:scond-d}
\frac{\partial_{\mu}p}{\varrho+p}=-{\partial_{\mu }W}+\frac{\Omega \partial_{\mu}\ell}{1-\Omega \ell},\quad \ell\equiv \frac{L}{E},\quad W\equiv\ln V_{eff}(\ell)
\\&&\nonumber
 V_{eff}(\ell)=u_t= \pm\sqrt{\frac{g_{\phi t}^2-g_{tt} g_{\phi \phi}}{g_{\phi \phi}+2 \ell g_{\phi t} +\ell^2g_{tt}}}, \quad
\mbox{where on} \quad\theta=\pi/2\\
&&
% \\&&\nonumber
%V_{eff}(\ell)^2={\frac{\Delta \rho ^2 \sigma ^2}{\sigma ^2 \left[a^4+a^2 \left(\ell^2+2 r^2\right)+r(r^3-4 a \ell M)\right]-\Delta \left(a^2 \sigma ^4+\ell^2\right)}}
%\\
%&&\mbox{where} \quad\sigma\equiv\sin \theta
V_{eff}^2(\ell)={\frac{\Delta \rho ^2}{ a^4+a^2 \left(\ell^2+2 r^2\right)+r(r^3-4 a \ell M)-\Delta \left(a^2+\ell^2\right)}}
\eea
(note that the effective potential $V_{eff}^2(\ell)$ is dimensionless). This potential is regulated by the radial profile of the geodesic specific angular momentum:
\bea
 \ell=\frac{a^3M +aMr(3r-4M)\pm\sqrt{Mr^3 \left[a^2+(r-2M)r\right]^2}}{[Ma^2-(r-2M)^2r]M}
\eea
}
while the  continuity equation  in Eq.\il(\ref{E:1a0})
is  identically satisfied as consequence of the applied conditions and symmetries.
{We note that  Eq.\il(\ref{Eq:scond-d}) is  a rearrangement of the Euler equation in Eq.\il(\ref{E:1a0}). We singled out
the variation of the specific angular momentum $(\ell)$   from the first term in \texttt{{r.h.s}} of Eq.\il(\ref{Eq:scond-d}), defining the effective potential and
 enclosing
the information on the gravitational component  of the torus forces balance.
A relevant aspect of Eq.\il(\ref{Eq:scond-d})  is the  shape of the toroidal configurations that can be found as the associated  exact integrals.
This is  possible due to symmetries of the system  and the assumption  of a barotropic equation  of state $p=p(\varrho)$.   In  this model, the fluid flow is not iso-entropic,  the entropy $ S $ is not constant in space and time but it is constant along the fluid flow, i.e. $u^a \nabla_a S=0$. In the  $V_{eff}$  definition,  the normalization condition for the fluid four-velocity, $u^au_a=-1$,  has been   taken into account, together with the definition of the parameterized specific angular momentum $\ell$. Assuming  $\ell=$constant, the second term of  \texttt{{r.h.s}} of  Eq.\il(\ref{Eq:scond-d}) vanishes.
This special re-writing of the forces balance  allows also identification of the critical points of the pressure. Both these aspects make the model extremely effective and versatile. On the other hand, there are several  possible generalizations of  this set-up: \textbf{1. }the first we mention here  concerns the inclusion of other components in the balance of forces. The   perfect fluid  energy momentum tensor (\ref{E:Tm}) can  include  for example a magnetic field component\cite{Zanotti:2014haa,Pugliese:2018zlx}. \textbf{2. }secondly, we might consider the influence of   the last term of Eq.\il(\ref{Eq:scond-d})   by adopting a varying fluid angular momentum  $\ell(r)$ as, for example, in \cite{Lei:2008ui,Witzany:2017zrx}.}

{Notice that we  singled out definition of specific angular momentum of the fluid $\ell\equiv E/L$ in terms of the functions $E$ and $L$, introduced in Eq.\il(\ref{Eq:after}); these quantities are constants of motion for test particle circular orbits, according to the presence of the two Killing fields $\xi_t$ and $\xi_{\phi}$, and in the test particle scenario in the BL coordinate frame they have an immediate interpretation   as quantities measured by  a static observer  at infinity. Nevertheless, for orbiting \emph{extended} matter  configurations, $(E, L)$ are not (generally) constants of motion and an interpretation of these quantities  has to be properly given.
Concerning the choice of $\ell=$constant, it is clear that in Eq.\il(\ref{E:Tm}) we considered a simple fluid, i.e., made up by one species particles, which is  subjected to pressure forces according to a barotropic equation of state. Other forces may have to be included as the  magnetic  or electric  components  in the energy momentum tensor. For the detailed discussion on relation between  $L$, $\ell$ and $E$ in the tori construction see for example \cite{pugtot}.}

We have therefore introduced the effective potential function    $V_{eff}(\ell)$  for the fluid
 %and the function $W$ known as  Paczynski-Wiita  (Paczynski) potential,
 which reflects the contribution of the background  Kerr geometry and the centrifugal effects. $\Omega$ is the relativistic angular frequency of the orbiting fluid relative to the distant observer. The  specific angular momentum $\ell$   is considered here constant and conserved (see also \cite{Lei:2008ui,Abramowicz:2008bk}).

As for the case of the  test particle dynamics, due to  the problem symmetries  we can limit the analysis to  positive values of $a>0$,
for \emph{corotating}  $(\ell>0)$ and \emph{counterrotating}   $(\ell<0)$ fluids and    we adopt the notation $(\pm)$  for  counterrotating or corotating matter  respectively
  (as $V_{eff}(\ell)$  in Eq.\il(\ref{Eq:scond-d})  is invariant under the mutual transformation of  the parameters
$(a,\ell)\rightarrow(-a,-\ell)$).

\subsection{Ringed accretion disks}\label{Sec:RADs}

We  consider a fully general relativistic model of ringed accretion disk \textbf{(RADs)} made  by several  corotating and counterrotating toroidal rings orbiting a supermassive Kerr attractor \cite{ringed}.
General Relativity
   hydrodynamic Boyer  condition  of equilibrium configurations  of rotating perfect fluids  governs the single torus.
   As a consequence of this many properties of the orbiting tori  are determined  by the  effective potential.
The
  toroidal surfaces  are the equipotential   surfaces of the effective potential  (and equipressure surfaces) $V_{eff}(\ell)$, considered  as function of $r$,  solutions of  $V_{eff}=K=$constant  or $ \ln(V_{eff})=\rm{c}=\rm{constant}$   \cite{Boy:1965:PCPS:}.  These  correspond also to the surfaces of constant density, specific angular momentum $\ell$, and constant  relativistic angular frequency  $\Omega$, where $\Omega=\Omega(\ell)$  as a consequence of the von Zeipel theorem \footnote{More generally $\Sigma_{\mathbf{Q}}$ is the  surface $\mathbf{Q}=$constant for any quantity or set of quantities $\mathbf{Q}$.  Therefore in this case $\Sigma_{i}=$constant for \(i\in(p,\rho, \ell, \Omega) \),  where the angular frequency  is indeed $\Omega=\Omega(\ell)$ and $\Sigma_i=\Sigma_{j}$ for \({i, j}\in(p,\rho, \ell, \Omega) \)} \cite{M.A.Abramowicz,Zanotti:2014haa}.

{Each Boyer surface turns to be  identified by the couple of parameters $(\ell,K)$.}
\subsection{RADs morphology}\label{Sec:RADmorpho}
A torus in the \textbf{RAD} agglomerate  can be corotating,  $\ell a>0 $, or counterrotating,   $\ell a<0$, with respect to the central black hole  rotation $a>0$. Consequently,  given a couple $(C_a, C_b)$ with specific  angular momentum $(\ell_a, \ell_b)$,  orbiting  in   the equatorial plane of a central Kerr \textbf{SMBH},   we can  introduce   the concept  of
 \emph{$\ell$corotating} tori,  defined by  the condition $\ell_{a}\ell_{b}>0$, and \emph{$\ell$counterrotating} tori defined  by the relations   $\ell_{a}\ell_{b}<0$. In other words,   a couple of  $\ell$corotating tori  can be both corotating $\ell a>0$ or counterrotating  $\ell a<0$ with respect to the central attractor\footnote{
In the following we will adopt often, when not required to do otherwise,
the  notation   which does not explicit the fluid  sign rotation. Then  the discussion    is intended to be  independent from this and  each ordering  relation must be understood for each $\ell$corotating sequence.}
{On the other hand , \emph{$\ell$counterrotating} couples are made of a  corotating torus  and  a counterrotating torus,  consequently the following two cases can occur:  a couple can be composed by an inner  corotating  torus,  with the respect to the $\mathbf{BH}$  and an outer counterrotating torus, or viceversa, the inner torus can be counterrotating and the  outer one corotating. }

\begin{figure}[h!]
\begin{center}
\begin{tabular}{lr}
\includegraphics[scale=0.3]{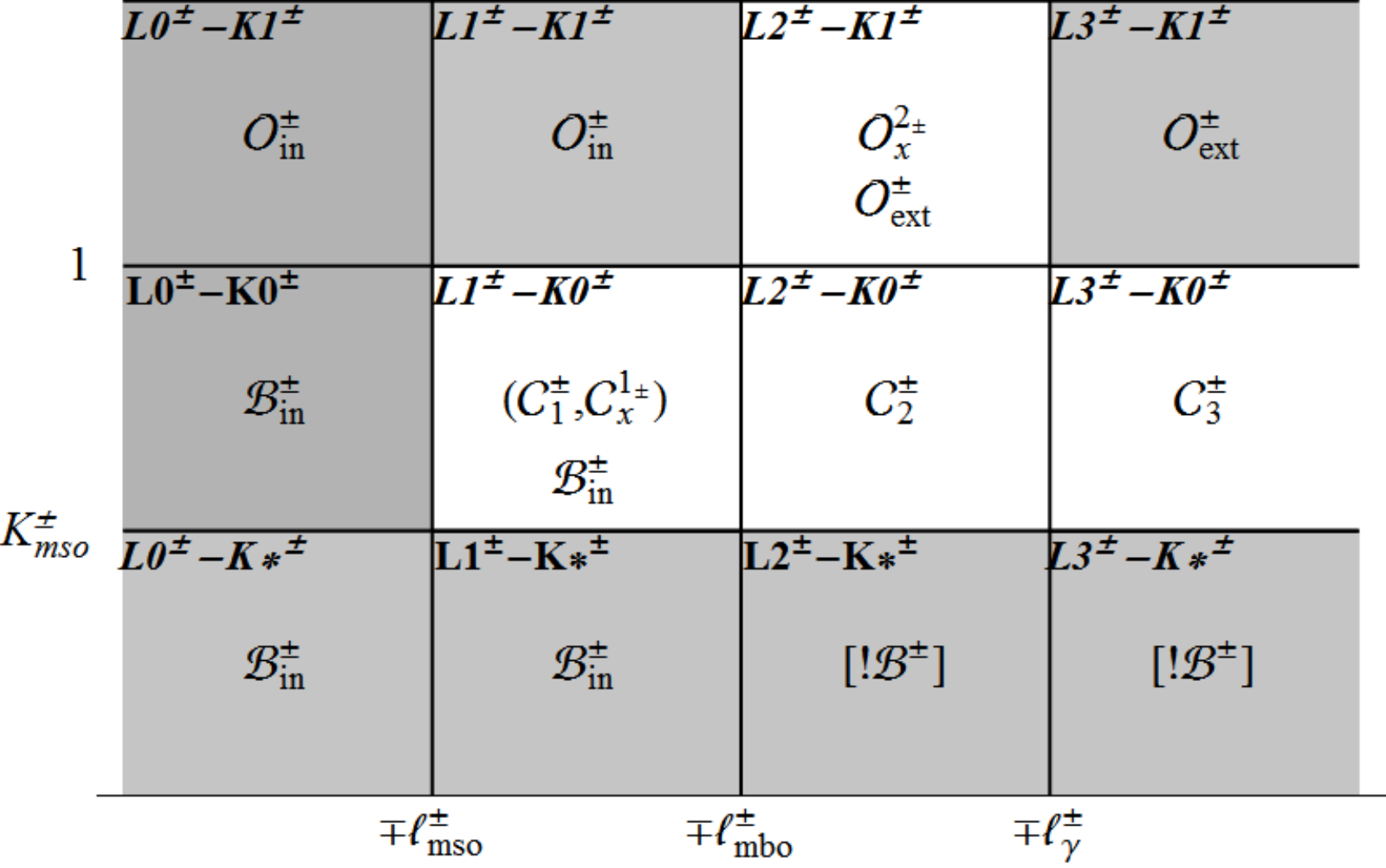}
\end{tabular}
\caption{Scheme of the variation ranges  $K\in \mathbf{Kj}$ where $j\in\{*,0,1\}$ and $\ell_i\in \mathbf{Li}$ where $i\in\{0,1,2,3\}$ for the corotating  $(-)$ and counterrotating $(+)$ cases respectively.
 $K_{mso}^{\pm}$ is the value of the parameters $K^{\pm}$ on the marginally stable orbit  $r_{mso}^{\pm}$,  while $\ell_{mso}^{\pm}\equiv \ell_{\pm}(r_{mso}^{\pm})$,  $\ell_{\gamma}^{\pm}\equiv \ell_{\pm}(r_{\gamma}^{\pm})$, $\ell_{mbo}^{\pm}\equiv \ell_{\pm}(r_{mbo}^{\pm})$ respectively. Radius $r_{\gamma}^{\pm}$ is the photon orbit, and  $r_{mbo}^{\pm}$ is  the marginally bounded orbit.
White regions indicate the ranges $\mathbf{Li-Kj}$  where critical points of the hydrostatic pressure are allowed.
The topology of the Boyer surface is also indicated: $C^{\pm}$ are the    closed regular toroidal surface  in equilibrium.
$C_x^{\pm}$ are the closed cusped surfaces  with accretion point.
$O_x^{\pm}$ are open surfaces   with a cusp (proto-jets) associated to the jet launch. The surfaces  $B^{\pm}$ and $O^{\pm}$ are not associated to the critical points of the pressure. {Sequences with $B^{\pm}$ and $O^{\pm}$ configurations for increasing $K$, with $\ell^{\pm}>\mp\ell^{\pm}_{mbo}$ constant  are more  articulated. In $\mathbf{L3-Ki}$ no $B$ surfaces appear $(\mathbf{[!B]})$. In $\mathbf{L2-Ki}$ the $B$ surfaces  are  associated also with values  $V_{eff}>1$, thus these configuration could also be as  $B_{in}$ or $B_{ext}$ kinds. Surfaces not associated to the critical points of the effective potential are  discussed of Sec.\il(\ref{Sec:lambda-right}). }--see  Figs\il\ref{Figs:LS01POLOt},\ref{Figs:ThinkPi}, a general  discussion can is in  Sec.\il(\ref{Sec:lambda-right})  and also \cite{pugtot}. The  superscript, or equivalently subscript $\{\mathbf{Q}_i, \mathbf{Q}^i\}$ with $i\in\{1,2,3\}$ respectively,  is for any quantity  $\mathbf{Q}$ relative to the range  of specific angular momentum $\mathbf{Li}$.}\label{Figs:Plotcredregrelre}
\end{center}
\end{figure}

Following this setup we  focus on the solution of Eq.\il(\ref{Eq:scond-d}), $W=$constant, associated to the critical points, {i.e., the extrema of the effective potential as functions of $r/M$, thus the minimum and maximum points of the effective potential, with angular momentum and parameter $K$ in the ranges, $(\mathbf{K0},\mathbf{Li})$ with $i=\{1,2,3\}$, and $(\mathbf{Kj},\mathbf{L2})$, $j=\{0,1\}$, represented  in Fig.\il\ref{Figs:Plotcredregrelre}. In fact, only for $K$ and $\ell$ parameters in these ranges, the effective potential function has    extreme points. Specifically  there are minima   for $K\in \mathbf{K0}$ and $\ell \in \{\mathbf{L1},\mathbf{L2},\mathbf{L3}\}$,  and there are maxima for $K\in \mathbf{K0}$ and $\ell \in \mathbf{L1}$, or $K\in \mathbf{K1}$ and $\ell\in \mathbf{L2} $.}
In Sec.\il(\ref{Sec:lambda-right}) we  briefly discuss  the  solutions  of   $W=$constant which are not  associated  to the extreme  points of the effective potential, therefore we shall consider the other   parameter regions of Fig.\il\ref{Figs:Plotcredregrelre} --see for details \cite{pugtot}.

Thus, more specifically we explore  the orbital region $\Delta r_{crit}\equiv[r_{Max}, r_{min}]$, whose boundaries correspond to the  maximum and minimum points of the effective potential respectively.
The inner edge  of the Boyer surface, the torus, must be   at $r_{in}\in\Delta r_{crit}$,
while the  outer edge of the torus  is at $r_{out}>r_{min}$.  Then, there is a further  matter configuration, which is    closest to the central black hole and it is located   at $r_{in}<r_{max}$.
 The limiting case of $K_{\pm}=K_{min}^{\pm}$ corresponds to a one-dimensional ring of matter  located in  $r_{min}^{\pm}$.

The centers, $r_{cent}$,  of the closed configurations $C_{\pm}$, where the hydrostatic pressure is maximum, are located at the minimum points  $r_{min}>r_{mso}^{\pm}$  of the effective potential. The toroidal surfaces are characterized by parameters $K_{\pm}\in [K^{\pm}_{min}, K^{\pm}_{Max}[ \subset]K_{mso}^{\pm},1[\equiv \mathbf{K0}$ and  specific angular momentum $\ell_{\pm}\lessgtr\ell_{mso}^{\pm}\lessgtr0$  for counterrotating and corotating fluids  respectively.

Therefore, the configurations have critical points, correspondent to the minimum and maximum of the hydrostatic pressure, more specifically then the maximum points of the effective potential $r_{Max}$  correspond to minimum points of the hydrostatic pressure and the  points  of  gravitational and hydrostatic instability.
An  accretion  overflow of matter from the  closed, cusped  configurations in   $C^{\pm}_x$ (see Figs\il\ref{Figs:LS01POLOt},\ref{Figs:ThinkPi}) can occur from the instability point  $r^{\pm}_x\equiv r_{Max}\in]r_{mbo}^{\pm},r_{mso}^{\pm}[$ towards the attractor, if $K_{Max}\in \mathbf{K0}^{\pm}$  with proper angular momentum $\ell\in]\ell_{mbo}^+,\ell_{mso}^+[\cup]\ell_{mso}^-,\ell_{mbo}^-[$ ($\mathbf{L1}^+\cup \mathbf{L1}^-$), respectively,  for counterrotating or corotating matter. Otherwise,  there can be  funnels of  material, associated to matter jets, along an open configuration   $O^{\pm}_x$ having    $K^{\pm}_{Max}\geq1$ ($\mathbf{K1}^{\pm}$), launched from the point $r^{\pm}_{J}\equiv r_{Max}\in]r_{\gamma}^{\pm},r_{{mbo}}^{\pm}]$ with proper angular momentum $\ell\in ]\ell_{\gamma}^+,\ell_{mbo}^+[\cup]\ell_{mbo}^-,
  \ell_{\gamma}^-[$ ($\mathbf{L2}^+\cup \mathbf{L2}^-$).
We could refer to the surfaces  $O^{\pm}_x$ as proto-jets, or for brevity jets\footnote{{ The role of ``proto-jet'' configurations,  corresponding   to limiting topologies for the  closed or closed cusped solutions associated with equilibrium or accretion,  is still  under investigation. More generally, in this model  the open surfaces  with $\ell\in\mathbf{Li}$  have been    always associated with the jet emission along the attractor symmetry axis--see for a general discussion \cite{KJA78,AJS78,Sadowski:2015jaa,Lasota:2015bii,Lyutikov(2009),Madau(1988),Sikora(1981)}.
 Although in Sec.\il(\ref{Sec:lambda-right})  we will briefly discuss  the more general open configurations, in this work we mainly analyze the cusped open configurations  considered as limiting surfaces associated with the two critical points of the   effective potential, the inner one where the hydrostatic pressure  is minimum, which is the instability point,  and the outer one  where the hydrostatic pressure is maximum, which is defined as  ``center'' of the open configuration. The Boyer surfaces with  sufficiently high specific angular momentum  and elongation, i.e. $(\ell, K)\in \mathbf{L2}\times \mathbf{K1}$ (corresponding to  centers which are far enough  from the attractor,  sufficiently high centrifugal component of the potential and sufficiently high density) are not capable to close  forming an  outer  disk edge, but there is still a point of instability, $r_J$, at the inner edge. Since  the nature of these configurations is a  limiting evolutionary stage  with respect to the other open topologies as well as to the  equilibrium disk or  closed disk  in accretion,  we referred to them shortly as proto-jet. It is clear that the problem to interpret these  configurations  invests  the more general problem of the  accretion-jet correlation  which  we also address in this work.  In terms of proto-jet configurations  ($\ell\in \mathbf{L2}$) it is  necessary to investigate the relation between jets emission and accretion, particularly with respect to the location of the inner edge of the accreting  disk (involving transition from the  instability  regions, for the parameter ranges  $(\mathbf{L1}, \mathbf{L2})$), the  magnitude of the disk specific angular momentum , and disk extension ($K\in \mathbf{K0}$ here also it linked to the size of the disk and its density and enthalpy). In this work, we attack the problem for the first time in framework of the  ringed accretion disk, while a more accurate analysis of the interpretation of these  open solutions, as well as of the emergence of other open  configurations  is left for further investigations.}}.

A second class of solutions are the  equilibrium, not-accreting, closed  configurations, with topology $C$,  defining  for $\pm\ell_{\mp}>\pm\ell_{mso}^{\mp}$ and centered in  $r>r_{mso}^{\mp}$ respectively. However there are  no  maximum points of the effective potential  for specific angular momentum $\pm\ell_{\mp}>\ell_{\gamma}^{\pm}$ ($\mathbf{L3}^{\mp}$), and therefore,  only equilibrium configurations are possible for fluids having specific  angular momenta in these ranges --see Figs\il\ref{Figs:LS01POLOt},\ref{Figs:ThinkPi}.

To simplify our discussion in the following   we use label $(i)$ with $i\in\{1,2,3\}$ respectively, for  any  quantity $\mathbf{Q}$ relative to the range  of specific angular momentum $\mathbf{Li}$ respectively, thus for example
$C_2^+$ indicates a closed  regular counterrotating configuration with specific angular momentum  $\ell_2^+\in\mathbf{L2}^+$.
For the \emph{ordered} sequences  of surfaces in the \textbf{RADs}, with the notation $<$ or $>$,  we intend  the ordered sequence of  maximum points of the pressure or $r_{min}$, minimum of the effective potential and the disk centers for the closed sub-configurations \cite{open}. In relation to a couple  of rings,   the terms ``internal'' (equivalently inner) or ``external'' (equivalently outer), will always refer, unless otherwise specified,  to the   sequence ordered according to the location of the  centers.
Then,  if   ${C}_{i}<{C}_j$ for  $i<j$,  ${C}_i$ is the inner ring, closest to attractor, with  respect to ${C}_j$, and  there is    $r_{cent}^i\equiv r_{min}^{i}<r_{min}^j\equiv r_{cent}^j$. Within these definitions, the rings  $(C_i, C_{i+1})$ and $(C_{i-1}, C_{i})$ are \emph{consecutive} as  $C_{i-1}<C_i<C_{i+1}$ \cite{ringed}.
The
symbols $\succ$ and $\prec$  refer instead to the sequentiality between the   ordered location of the \emph{minimum} points of the pressure,   or $r_{Max}$, maximum point  of the effective potential, if they exist, which are  the instability points of accretion, $r_x=r_{Max}$ for $C_x$ topologies,  or of launching of jets, $r_{Max}=r_J$ for the open cusped topologies (proto-jets).
Therefore it is always
$r_{min}^i<r_{min}^o$.  We note that only for an $\ell$corotanting sequence this definition  \textit{implies} also   $r_{Max}^i>r_{Max}^o$. In other words,  for the $\ell$corotating sequences there is  always $()_i<()_o$ and $()_i\succ()_o$, whereas for an $\ell$counterrotating couple this is not  always verified \cite{open,dsystem,Long}. Here $()$ means  any closed or open configuration.

As for the  sequences of open configurations, $O_x$, we shall  mainly deal   with  the  relations between the  critical points of the configurations, then it will be  convenient  to introduce  the \emph{criticality indices}  $\hat{i}$, which are  associated   univocally to the couple ($r_{Max}^i$, $\ell_i$), where $i$ is as usual the configuration index   univocally associated to ($r_{min}^i$, $\ell_i$).
We can note that
 $\hat{i}(i)$ is a  decreasing function of the configuration index $i$ (ordering the maximum of pressures) or  $\partial_i \hat{i}(i)<0$. Thus, for example,
$O_x^i\succ O_x^o$ and  $O_x^{\hat{o}}\succ O_x^{\hat{i}}$   where we have $|\ell_i|=|\ell_{\hat{o}}|<|\ell_o|=|\ell_{\hat{i}}|$.

\begin{figure}[h!]
\begin{center}
\begin{tabular}{lr}
\includegraphics[scale=0.2]{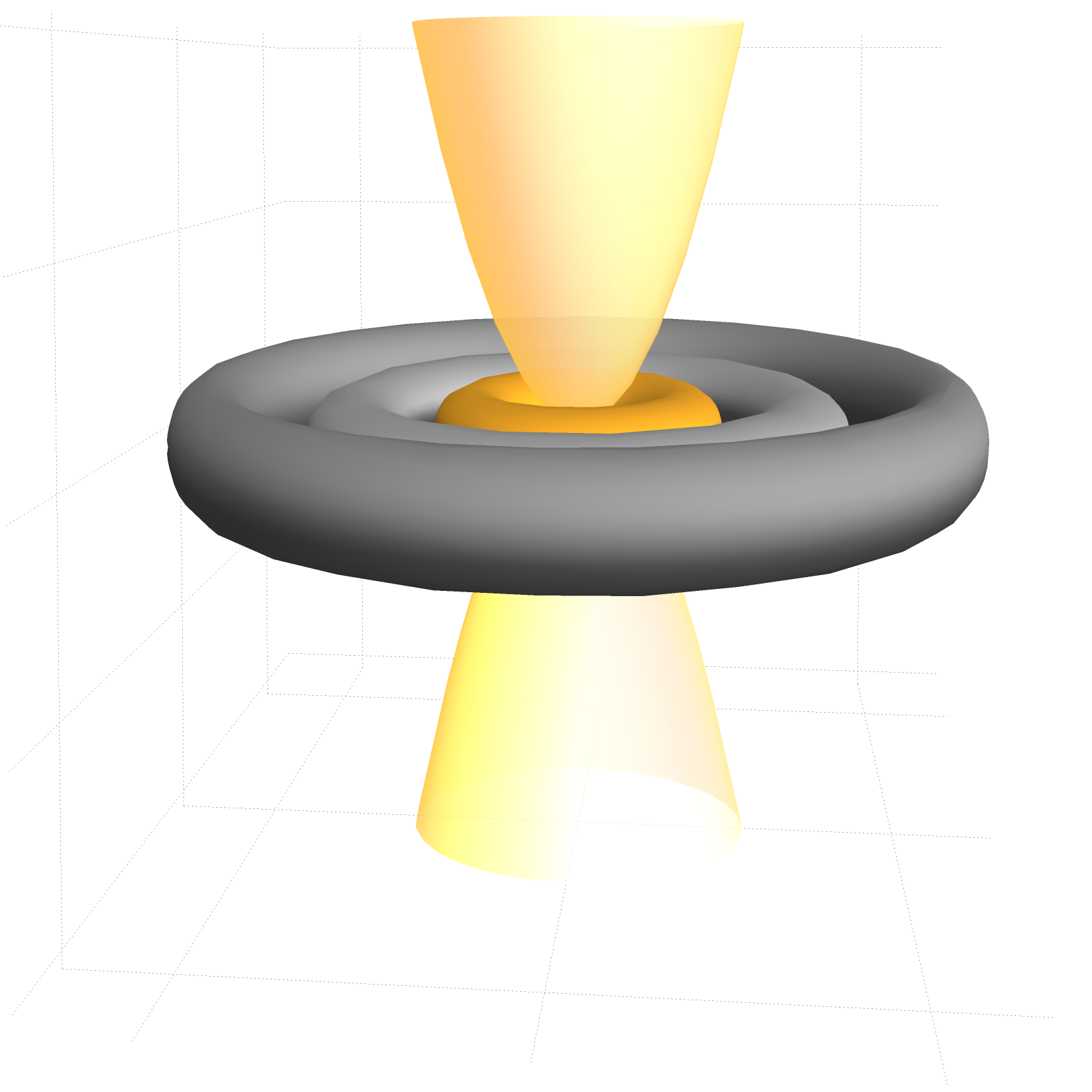}\\
\end{tabular}
\caption{Pictorial representation of a ringed accretion disk with  open $O_x$ surfaces.}\label{Figs:Plotcredregrelre3D}
\end{center}
\end{figure}
{Finally, it should be noted that
the simple case of barotropic tori with constant distribution of specific angular momentum provides all the relevant details of the  \textbf{RAD} structure, as made up by thick toroidal disks, and at the present level of our knowledge it is a proper approximation to more realistic models.
The effects on the adoption of more general laws  for the fluid specific angular momentum  in the tori model has been extensively discussed  in the   literature--see for example \cite{Lei:2008ui,Witzany:2017zrx}.
Then, concerning the  ``piecewise'' structure  of the distribution of  the ring specific angular momenta (the differential rotation of the ringed disk \cite{ringed}),  the  rings  are supposed to be  formed  in different accretion regimes  and therefore it  is  natural (and necessary for the tori separation) that  each torus is    characterized by a different specific angular momentum.}
\section{RADs instabilities}\label{Sec:insta-Rad-Proc}
The existence of a  minimum of the   hydrostatic pressure ($r_{Max}$)   implies  the existence of a  critical topology for the fluid configuration--Figs\il\ref{Figs:Plotcredregrelre},\ref{Figs:LS01POLOt},\ref{Figs:ThinkPi}.
%%%%%%%%%%%%%%%%%%%%%%%%%%%%%%%%%%%%%%%%%%%%%%%%%%%%%%%%%%%%%%%%%%%%%%%%%%%%%%%%
To sum up the situation for a \textbf{RAD} we may say there are  two classes of    points  associated to the instability  of the  macro-structures: \textbf{{1.}} the  Paczynski instability points (corresponding to violation of mechanical equilibrium in orbiting fluids see also \cite{Paczynski:2000tz,AbraFra}), maxima of the effective potential   $r_{Max}\in\{r_x, r_J \}$ for a $O_x$ or a $C_x$ topology and  \textbf{{2. }}  the contact points $r^i_{\odot}$  (see \cite{ringed}) between two surfaces of the decomposition featuring tori collision and   associated to the second  mode of instability  for a $\mathbf{C}_{\odot}$ ringed disk. Each   ring may admit a maximum of two contact points    $r^i_{\odot}$
\footnote{The cusps $(r_x, r_J)$ of  the  corotating  surfaces can also occur  in the ergoregion in the spacetimes with $a>a_1$, while for spin $a=a_1\equiv1/\sqrt{2}M\approx0.707107M$ the unstable point  is on the static limit--see \cite{ergon} and also \cite{Bejger:2012yb} and Fig.\il\ref{Figs:LS01POLOt}. These surfaces are detailed in Sec.\il(\ref{Sec:lambda-right}).}.

Consequently, we can  identify two successive instability phases  of the macro-configuration: the {first} \textbf{(I)} with  the formation of one or more  points of instability or involve two sub-configurations in the case of formation of a macro-configuration $\mathbf{C}_{\odot}$. \textbf{(II)} The second phase consists  in the global instability of the ringed disk  which    follows  the first phase. For both instability modes, in the second phase a penetration of matter from one unit to  another is expected, and this can   result in   possible destabilization of the entire macro-structure by collisions between fluids.
Then, although the second phase  in the two modes may be similar, the essential  difference is in the first phase of instability which is, in the first mode  induced by an instability point, while in the second mode,  the presence  of contact points is necessary.
The Paczynski mechanism in the first mode may be   involved as the cause of the second phase, while  in the second mode, starting with a $\mathbf{C}_{\odot}$ topology, of  Paczynski  instability  could emerge  in the second phase as a  possible effect of the first phase of instability.
 In this work, we analyze the situation only in the first phase.
%%%%%%%%%%%%%%%%%%%%%%%%%%%%%%%
%================================
%
\subsection{Geometrical correlations of unstable configurations}\label{Sec:geometrical-corre}
Concerning the possibility of  collision and emergence of instability we introduce the concept of geometrical and causal correlation for unstable configurations of the \textbf{RAD}.
Two sub-configurations of a ringed disk are said to be geometrically correlated if they are in contact, or they may be in contact according to some constraints settled on their morphological or topological evolution and the geodesic structure of the spacetime, i.e. according to the effective potentials they are subjected to \cite{ringed}.
{We note that, since the intersection of the  geodesic structures defined by $r_{\mathcal{N}}^{\pm}$, and introduced in Sec.\il(\ref{Sec:models}),  is not empty, the analysis of the geodesic structure  will be particularly  important for the characterization of the  geometrical correlation for    the $\ell$counterrotating sequences and particularly for the mixed subsequences.}
Thus part of this analysis is devoted to figure out whether and when two sub-configurations of a decomposition can be in contact, or, in fact, geometrically correlated according  to a number of features set in advance, or which  we seek to establish. A contact in this model causes  collision and  penetration of matter, eventually with the feeding of one sub-configuration with material and supply of specific angular momentum of  another consecutive  ring of the decomposition. This mechanism  could possibly end  in a change of   the ring disk  morphology and topology.  In fact, several instability points can be  geometrically correlated and present in a first phase of  macro-structure instability,  and being causally correlated,  arising  in a second phase of the macro-structure instability.
 The existence of a contact point in this model is  governed by the   geodesic structure of the Kerr geometry.
It is clear that a geometrical correlation in a ringed structure  induces a  causal correlation in a couple, when
the morphology or topology of an element can be regarded as a result of that correlation,
which arises by the geometry  of the common attractor\footnote{We note that
the problem of inferring and even to define a causal correlation between events or    objects  can be  indeed subtle and it is certainly  relevant in a  variety of scientific disciplines--for a very general  discussion we  refer to \cite{quantum}.    In the case of jet-proto-jet correlation  and the jet-accretion correlation proposed  here we should consider that generally correlation   does not necessary imply causation,  and this  latter aspect of the  correlation should be  faced more deeply together with a more general discussion of the correlation  definition used here. }.

%%%%%%%%%%%%%%%%%%%%%%%%%%%%%%%%%%%%%%%%%%%%%%%%%%%%%%%%%%%%%%%%%%%%%%%%%%%%%
\subsection{Parameter setting, rotational symmetry and  $\breve{\ell}$}\label{Sec:div-spee}
In this section we investigate some symmetric properties of  the angular momenta
$\breve{\ell}:\; V_{eff}(\ell,r)=1 $  introduced  throughout  the discussion of \cite{open,Long}, for a general radius $r$ on the equatorial plane.
This special specific angular momentum   allows us to discuss some properties of symmetry of the effective potential inherited from the background symmetries and the toroidal motion of the orbiting  system.
An important role of the momentum  $\breve{\ell}$  has been widely studied as $\ell_{mbo}: V_{eff}(\ell_{mbo},r_{mbo})\equiv K_{mbo}=1$ and $\partial_r   V_{eff}=0$ in $r=r_{mbo}$. Then clearly this is the asymptotic value of the effective potential for  large $r/M$  (or large $R=r/a$).

We can  then reduce this problem to find out the solutions of the following equations
\bea\label{Eq:anas}
&&
{2 \Delta_-^2-\left(a^2+\Delta_- \Delta_+\right) r+2 r^2}=0\quad\mbox{where}
\\
&&\nonumber V_{eff}(\Delta_{\pm},r;a)^2\equiv\frac{r \Delta }{2 \Delta_-^2-\Delta_- \Delta_+ r+r^3}>0,
\\
&&\label{Eq:simm-MalD-BH}
\Delta_{\pm}\equiv\ell\pm a,\quad \Delta_- \Delta_+ >0,\quad  \Delta_{\mathbf{Q}}\gtrless 0\quad\mbox{if}\quad  \ell\gtrless0,
\\
&&\nonumber \Delta_{\mathbf{Q}}=\{\Delta_+, \Delta_-\},\quad a>0,\quad |\ell/a|>1.
\eea
%
%\btb
{{For sake of simplicity  in this Section we  mainly use  dimensionless quantities where $r\rightarrow r/M$ and $a\rightarrow a/M$}.}
{Equation (\ref{Eq:anas}) is obtained  by the condition $V_{eff}^2-1=0$ on the equatorial plane $(\theta=\pi/2)$, while  $V_{eff}(\Delta_{\pm},r;a)^2$ is  a rearrangement  and a new parametrization of $V_{eff}(r;\ell,a)^2$ of Eq.\il(\ref{Eq:scond-d}) using
new $\Delta_{\pm}$. }
Note that  $\Delta_{\pm}(-\ell)\lessgtr0$ for $\ell\gtrless0$, and  the quantity  $\Delta_{\pm}$ in  Eq.\il(\ref{Eq:anas}) should not be confused   with $\Delta$ in  the metric given by  Eq.\il(\ref{alai}).
No critical points exist, therefore, no Boyer surface  exists, for $|\ell/a|<1$ \cite{pugtot}.
We will use  Eq.\il(\ref{Eq:simm-MalD-BH}), exploiting the symmetries in the couple $\Delta_{\pm}$ for sign reversal in the specific angular  momentum.
We have:
\bea\label{Eq:we-next}
&&\ell=\frac{\Delta_-+\Delta_+}{2},\quad a=\frac{(\Delta_+-\Delta_-)}{2},
\\\nonumber
&& V_{eff}(\Delta_{\pm},r)\equiv\frac{1}{2} \sqrt{\frac{r \left[(\Delta_--\Delta_+)^2+4 (r-2) r\right]}{2 \Delta_-^2-\Delta_- \Delta_+ r+r^3}}.
\eea
 Then we can introduce, a single rotation parameter  for the system accretor-disk rather than two,
defined by:
\be
\mathcal{A}_{\pm}\equiv \frac{\Delta_{\pm}\mp \Delta_{\mp}}{2}: \quad
\mathcal{A}_+= a>0, \quad \mathcal{A}_-=\ell,
\ee
%
%*******************************************
where  the two rotational parameters $\ell$ and $a/M$ are now replaced by the two $\Delta_{\pm}$.

The difference in the two forms of the potential in Eq.\il(\ref{Eq:anas}) and Eq.\il(\ref{Eq:we-next}), respectively. This definition  could be important in the analysis of accretion disks, because this  re-parametrization may result in the identification of significant traceable quantities, when the fluid specific angular  momentum neither  the  attractor spin are not made explicit.  Therefore, we expect these symmetries will be  deepened further in a future work.
 %+
Solving explicitly the problem (\ref{Eq:anas})  in terms of $\ell$, one gets the two solutions
\bea\nonumber&&
\breve{\mathfrak{l}}_{\mp}\equiv\mp\frac{\pm2 a+ \sqrt{2r \Delta }}{r-2},\quad \breve{\mathfrak{l}}_{-}<0 \quad \forall r,\quad \mathfrak{\breve{l}_{+}}<0\quad\mbox{in}\quad r<r_{\epsilon}^+,
\\&&\label{Eq:formula-ris}
\mbox{where}\quad\breve{\ell}_{\pm}=\breve{\mathfrak{l}}_{\pm}(r_{mso}^{\pm})\quad \breve{\ell}^-_{2_+}=\breve{\mathfrak{l}}_{+}(r_{mso}^-).
\eea
Therefore, the specific angular  momentum $\breve{\ell}$ is provided by    Eq.\il(\ref{Eq:formula-ris}) and, considered the form of the radii $r_{mso}^{\pm}$,
the  symmetries between of the $\ell$counterrotating  fluid configurations at  the marginally stable orbits  are clear.
However, we can  investigate more deeply this aspect, using the variables  $\Delta_{\pm}$.
Then the solutions  (\ref{Eq:anas}) are the couples:
\bea&&
(\Delta_-^{[-]},\Delta_+^{[+]})\quad\mbox{and}\quad( \Delta_-^{[+]},\Delta_+^{[-]})\quad\mbox{where}
\\\nonumber
&&\Delta_-^{[\pm]}\equiv-\frac{a r\pm\sqrt{2} \sqrt{r \Delta }}{r-2},\quad\Delta_+^{[\pm]}\quad\frac{a (r-4)\pm\sqrt{2} \sqrt{r \Delta }}{r-2}.
\eea
Considering  the following symmetries:
\bea\label{Eq:Simm-Ver-UE}
&&\Delta_+(\ell)=-\Delta_-(-\ell),\quad \Delta_+(-\ell)=-\Delta_-(\ell)
\\\label{Eq:stat-nt}
&&\Delta_-(-\ell)^2=\Delta_+(\ell)^2,\quad
\Delta_-(\ell)^2=\Delta_+(-\ell)^2,
\\
&&\nonumber
\Delta_+(\ell) \Delta_-(\ell)=\Delta_-(-\ell) \Delta_+(-\ell)
\eea
and having in mind also Eq.\il(\ref{Eq:simm-MalD-BH}),  we can say that  a change in sign $-1\Delta_{\mathbf{Q}}$ acts in exchanging   the subscript sign $\Delta_{\pm}$  and  $\ell$.
Then we can solve the Boyer problem  to find out the ring surfaces  in terms of the resolving couples
$\Delta_{\pm}$.   By considering  effective potential $V_{eff}(\Delta_{\pm},r)^2$ in Eq.\il(\ref{Eq:anas}), we get the following solutions for the specific angular  momenta of the critical points for the pressures where solutions of the Boyer problem exist:
\bea
&&
\Delta_-= \frac{a (r-1) r^2-\sqrt{r^3 \Delta ^2}}{\Delta +X}, \quad{\mbox{and}}
\\\nonumber
&&\Delta_+= -\frac{-\{2 a^3-a r [8+(r-7) r]\}+\sqrt{r^3 \Delta ^2}}{\Delta +X},
\\
&&
\Delta_-= \frac{a (r-1) r^2+\sqrt{r^3 \Delta ^2}}{\Delta +X} \quad{\mbox{and}}\\
\nonumber
&&\Delta_+= \frac{2 a^3-a r [8+ r (r-7)]+\sqrt{r^3 \Delta ^2}}{\Delta +X}, \quad X\equiv - r (r-2) (r-1).
\eea
where $\Delta_+-\Delta_-=2a$.
Similar considerations are also possible considering the situation on different planes, for example by considering the quantity   $\ell/a\sin \theta$, see for example \cite{pugtot}, and  possibly generalizing the   specific angular  momentum definition \cite{Lei:2008ui}.

However, it is convenient to take advantage of the symmetries of the configuration in the treatment of extended systems of matter in the  axissymmetric  fields,  and particularly  the  symmetry of  reflection on the equatorial plane.
In the re-parametrization  $(\ell,a/M)\mapsto \Delta_{\pm}$  we can take advantage of the symmetries in Eq.\il(\ref{Eq:Simm-Ver-UE}),  managing  only one sign, then reducing
\emph{to one only},  \emph{always positive}  variable $\Delta_+$ or $\Delta_-$, without considering
the corotation or counterrotation nature of the fluid (that could indeed be difficult to be assessed)
but considering Eq.\il(\ref{Eq:stat-nt}).
 The  square  $\Delta_-^2$  is the only term of  the potential  in Eq.\il(\ref{Eq:anas}) that changes subscript after a change of the  specific angular   momentum. Suppose for simplicity  $\ell>0$,
then, being $\Delta_{\pm}$,  the variable, we can solve for one of the two $\Delta_{\pm}$  and then use any  of the symmetries in Eq.\il(\ref{Eq:Simm-Ver-UE}) to infer information on the other one.
\section{Open equipotential surfaces and  fluid configurations}\label{Sec:lambda-right}
The existence of the fluid  configurations is schematically summarized in Fig.\il\ref{Figs:Plotcredregrelre} in terms on the ranges of the specific angular momentum and the $K$ parameter.
\begin{figure}[h!]
\begin{center}
\begin{tabular}{cc}
\includegraphics[width=.5\textwidth]{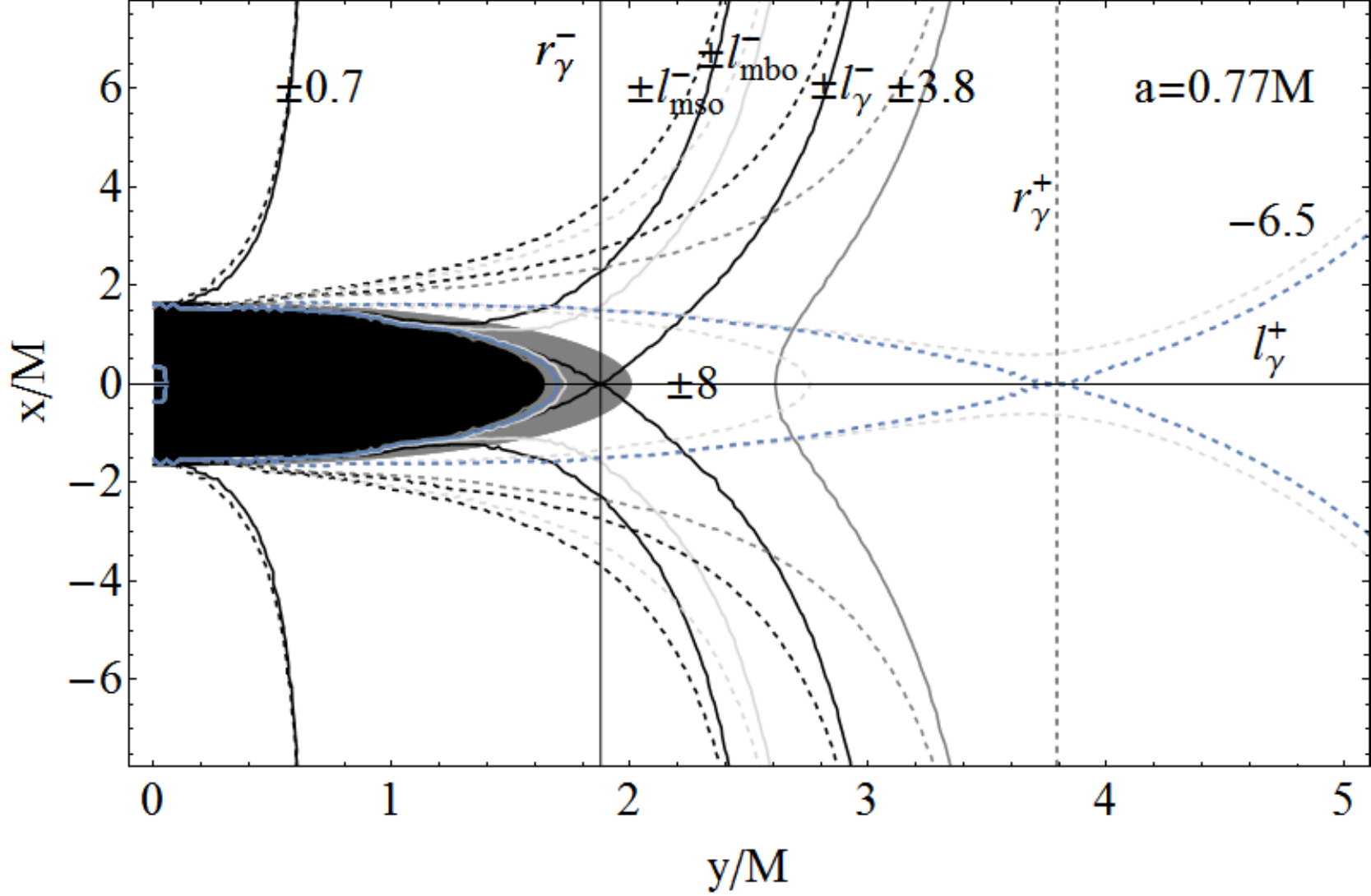}
\\
\includegraphics[width=.5\textwidth]{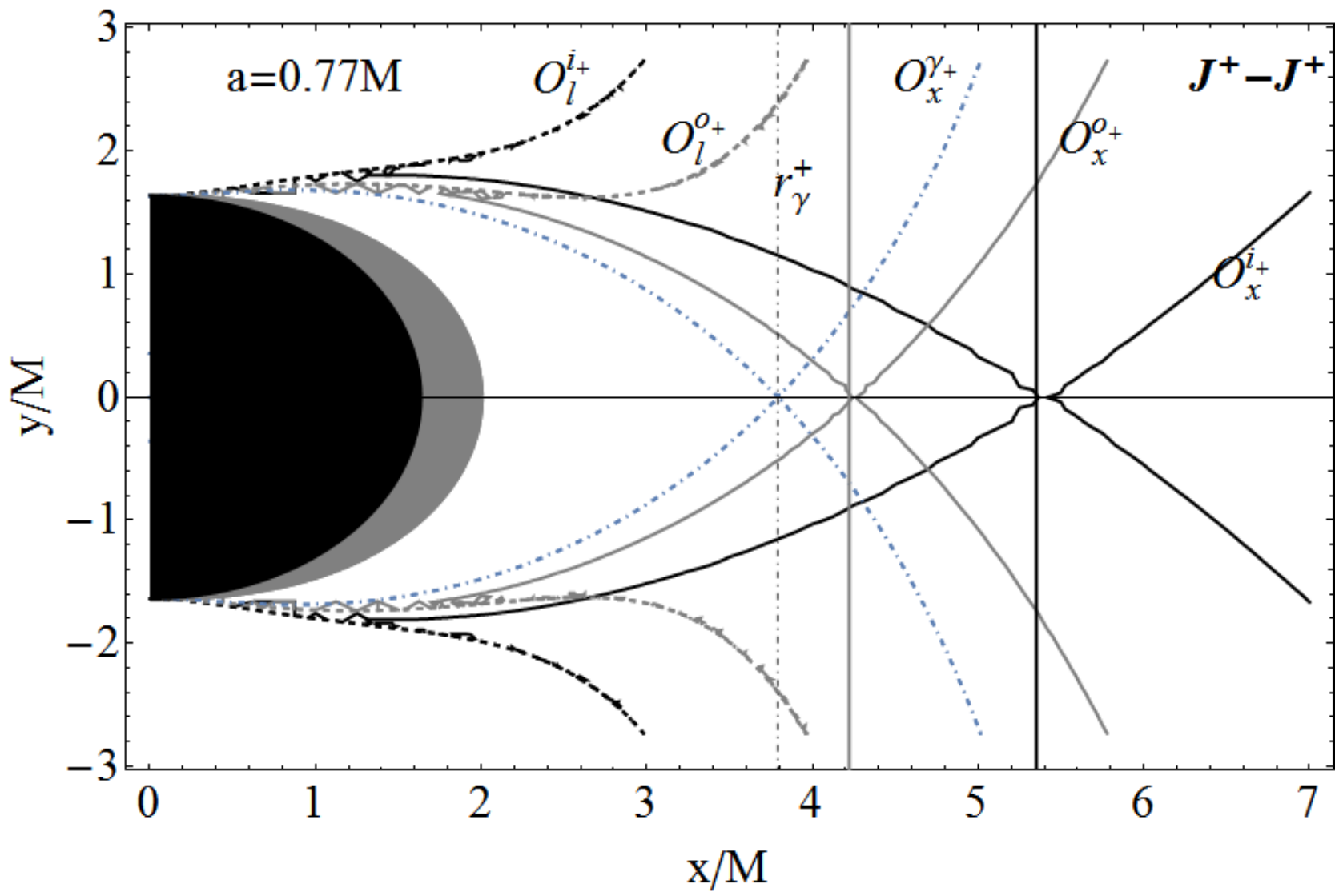}
\end{tabular}
\caption{\emph{Upper}: Spacetime spin $a=0.77M$, $\ell$counterrotating sequences, $\ell_i\ell_j<0$. The outer horizon is  at $r_h=1.63804M$, the region $r<r_h$ is  black-colored, the region $r<r_{\epsilon}^+$ is gray colored, radius $r_{\epsilon}^+$ marks the static limit. Surfaces in the $x-y$ plane, for different values of specific angular momentum $\ell$ for corotating ($\ell>0$-continuum lines) and counterrotating ($\ell<0$--dashed lines) fluids. $r_{\gamma}^{\pm}$ are the photon circular orbits were $\ell_{mso}=\ell(r_{mso})$ and $\ell_{\gamma}=\ell(r_{\gamma})$. Radius $r_{mso}$ is the marginally stable orbit. The superscript $()_{\pm}$ is for corotating $(-)$ and counterrotating $(+)$ matter respectively. The numbers  close to the curve are the values of the specific angular momentum \cite{open}. \emph{Below}:pacetime spin $a=0.77M$, $\ell$corotating sequences, $\ell_i\ell_j>0$, of counterrotating open configurations $\mathbf{J^+-J^+}$  $\ell_i a<0$ $\forall i j$. Decomposition including open-crossed sub-configurations ($\gamma$-surface) $O^{\gamma}_x$, open cusped with angular momentum $\ell_{\gamma}$.  The outer horizon is  at $r_h=1.63804M$, black region is $r<r_h$, gray region is $r<r_{\epsilon}^+$, where $r_{\epsilon}^+$ is the static limit, and $r_{\gamma}^+$ is the photon orbit on $\Sigma_{\pi/2}$. For $O^{o_{+}}_x$ there is  $\ell_o=-5.62551$ and $ K_o=1.28775$, for
$O^{i_+}_x$  there is $\ell_i=-4.66487$ and $ K_i=1.00272$. The $h\gamma$-surfaces, $O_{l}^{i_+}$ and $O_l^{o_+}$ are also plotted--see also Sec.\il(\ref{Sec:lambda-right})}\label{Figs:LS01POLOt}
\end{center}
\end{figure}
%
%%
%\begin{figure}[h!]
%
%
\begin{figure*}[h!]
\begin{center}
\begin{tabular}{cc}
\includegraphics[width=.6\textwidth,angle=90]{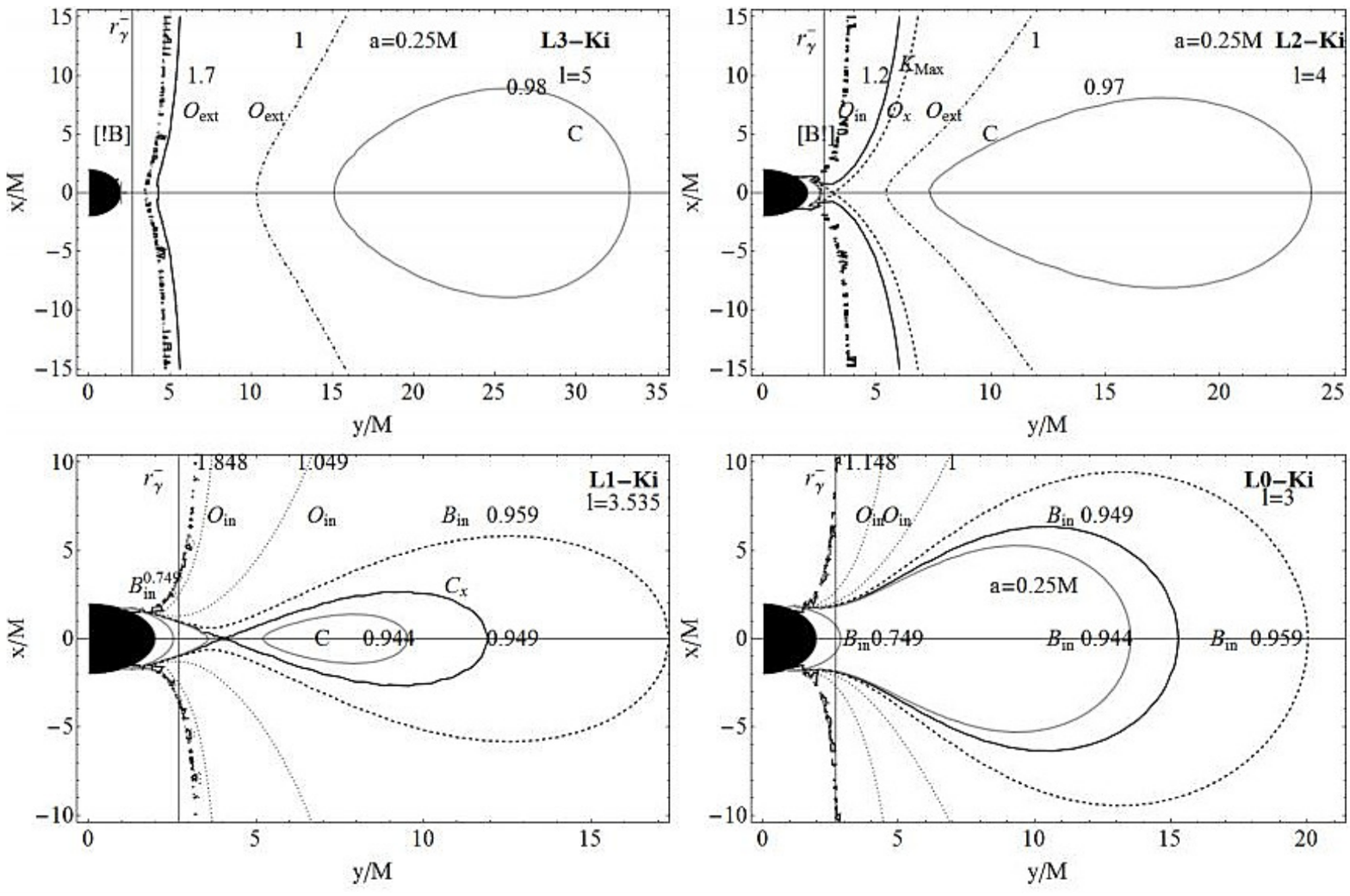}%
\end{tabular}
\caption{Spacetime spin $a=0.25M$, $\ell$corotating sequences, $\ell_i\ell_j>0$, of corotating disks  $\ell_i a>0$ $\forall i j$. Decompositions including open-crossed sub-configurations $O_x$.  The outer horizon is  at $r_h=1.96825M$, the region $r<r_h$ is  black-colored. In $\mathbf{L3-Ki}$ no $B$ surfaces appear $(\mathbf{[!B]})$. Surfaces not associated to the critical points of the effective potential are  discussed of Sec.\il(\ref{Sec:lambda-right}). }\label{Figs:ThinkPi}
\end{center}
\end{figure*}
In \cite{open}  we    discussed   the configurations associated with the pressure critical points, occurring for the parameter ranges   $\mathbf{Li}\cup \mathbf{Kj}$, with $\mathbf{Ki}\in\{\mathbf{K0},\mathbf{K1}\}$, and $\mathbf{Li}\in\{\mathbf{L1}, \mathbf{L2}\}$, or $\mathbf{L3}\cup \mathbf{K0}$.  Here, we add some comments on  the surfaces not associated to the   critical points of the effective potential function but the solutions of $V_{eff}(\ell,r)=K$.
In the following, we address the discussion   in terms  of the specific angular momentum  magnitude unless  the fluid rotation with respect to the black hole is not explicitly specified.

We start our considerations by noting that
the hydrostatic pressure has a  monotonic behavior as a function of $r/M$ for specific angular momentum in
$
\mathbf{L0}: \ell<\ell_{mso}$, where $V_{eff}<1$ on the equatorial plane.
For these values of specific angular momentum  $\ell$, there are are no critical points of the fluid pressure.
 As proved in   \cite{pugtot}, the case $\bar{\ell}\equiv |\ell/a| <1$ is a restriction of $\mathbf{L0}$ on the equatorial plane\footnote{
  On the  planes different from  the equatorial one,  we can generalize the limit by assuming ${\bar{\ell}}\equiv \ell/a\sigma <1\in]-1,1[$ or, as discussed below,  $\bar{\bar{\ell}}\equiv \ell/a\sigma \in]-1,1[$, with $\sigma=\sin \theta$, where no critical points are possible--see also \cite{pugtot}.}.
At $K<1$, range \textbf{K0}, and  $\ell<\ell_{mso}$, range \textbf{L0}, no critical points of the effective potential  occur on the equatorial plane: the pressure generally decreases with the radius (on the plane $\Sigma_{\epsilon}^{+}$).
However, at $\mathbf{L0}\cup\mathbf{K0}$ there are  the  $\mathbf{B}_{in}$  surfaces, fat closed tori
 as shown in Fig.\il\ref{Figs:ThinkPi}. Funnels  are not associated to these solutions.
 The unique  solution of the problem $V_{eff}=K<1$ corresponds to the  outer edge of this configurations. The surface becomes smaller for decreasing $K\in \mathbf{K0}$. The surface area increases by decreasing the specific angular  momentum  magnitude $\ell\in \mathbf{L0}$.
 Qualitatively  we could conclude that   the rotation relative to the attractor plays an irrelevant role in the determination of the morphology of these surfaces,  which confirms    a symmetry between the $\ell$counterrotating sequences as pointed out in \cite{pugtot} and \cite{ringed}.
 However,  these  features distinguish morphologically the  $\mathbf{B}_{in}$ from the lobe $\mathbf{B}_{ext}$ which appears  at higher specific angular momenta.
Decreasing the specific angular  momentum from  {{starting value}}  $\ell \in \mathbf{L3}$, when there is a maximum of the hydrostatic pressure,  a first surface  $B_{ext}$  occurs.  At $K\geq K_{min}$, a toroidal ring is formed, this  grows  and approaches the attractor as the specific angular  momentum magnitude decreases.
If  $K_{Max}<1$, i.e.  $\ell \in \mathbf{L1}$, then  with decreasing  of the specific angular   momentum  in $\mathbf{L1}$ towards values in $\mathbf{L0}$, a $B_{in}$  inner surface  appears (i.e. one has  the sequence of configurations $\{\mathbf{B}_{ext},\mathbf{C}, \mathbf{C}_x, \mathbf{B}_{in}\}$), with no open funnel--Fig.\il\ref{Figs:ThinkPi}.

 For
 $K\geq1$,  open surfaces of  funnels of matter  appear,  close to the rotation axis  as  the specific angular  momentum magnitude  decreases in $\ell\in \mathbf{L0}$. These configurations will be referred as $O_{in}$.
This surfaces (where  $\partial_x y>0$)  do not cross  the axes $y=0$--see Fig.\il\ref{Figs:ThinkPi}
\footnote{Increasing  $K\in \mathbf{K1}$,  or  decreasing $|\ell|\in \mathbf{L0}$  a  {collimation} occurs, i.e., there would be
$ \left.\partial_{|\ell|}{|y|}\right|_{x}<0$ but $\left.\partial_{|x|}{|y|}\right|_{|\ell|}>0
$.
As in \cite{Long}, we can define {collimation} of the funnels  along the rotation an  axis  if there is at lest one
$x_c:\;  \partial_{|x|} |y|< 0$ for $ x>x_c$
where the rotation axis is located at $y=0$.
The open surface will be said {collimated}, the coordinate $x_c$, and the corresponding $y_c$, give the collimation points.  If $ \partial_{|x|} |y|= 0$, in a given finite region of  $ x $, then the funnel structure is tubular, formed by  matter  rotating  around the $x$ axis at each  $\Sigma_t$ (in this model $\dot{r}=0$). In the tubular structure the radius  of the cylinder remains constant for each $y$.
In the proto-jets there should be some mechanism pushing the matter to
induce a $\partial_{|x|}|y|<0$ by changing also accordingly the  specific angular momentum magnitude  or $K$ parameter. For example, one might ask if a change of spin of the attractor  could constitute  such a factor, or  also  one can equally consider the case of the open critical configurations $O_x$. {For a general discussion on the role of the open surfaces and their connection with jet emission we refer to \cite{KJA78,AJS78,Sadowski:2015jaa,Lasota:2015bii,Lyutikov(2009),Madau(1988),Sikora(1981)}.
Then, given an $\ell$corotating $O_x^{\hat{i}}\prec O_x^{\hat{o}}$  couple, it is always
$O_x^{\hat{i}}\supset O_x^{\hat{o}}$ at  $r<r_J^{\hat{o}}$ and $O_x^{\hat{i}}\subset O_x^{\hat{o}}$ for $r>r_J^{\hat{i}}$, and there is  a couple of cross points ($O_x^{\hat{i}}\cap O_x^{\hat{o}}\neq0$) in $]r_J^{\hat{i}},r_J^{\hat{o}}[$--see Fig.\il\ref{Figs:LS01POLOt}. This means that,
given $\bar{x}<r_j^{\hat{o}}$ on the equatorial plane,
there is $|O_x^{\hat{i}}|\equiv |\bar{y}^{\hat{i}}|>
|O_x^{\hat{o}}|\equiv|\bar{y}^{\hat{o}}|$.
In other words, the curve  $O_x^{\hat{o}}$ is contained  in  the region of the $y-x$  plane  with boundary  $O_x^{\hat{i}}$ for $r
<r_J^{\hat{o}}$.
Viceversa, at $r>r_J^{\hat{i}}$  the (open) surface
$O_x^{\hat{i}}$  is contained in the region of the  plane cut by
 $O_x^{\hat{o}}$, or
 for  $\bar{x}>r_j^{\hat{i}}$ on the equatorial plane,
there is $|O_x^{\hat{i}}|\equiv|\bar{y}^{\hat{i}}|<
|O_x^{\hat{o}}|\equiv|\bar{y}^{\hat{o}}|$. This means, in particular, that the funnels  of  $\ell$corotating couple of open surfaces  do not cross each other  in the region  $r>r_J^{\hat{o}}$  and generally the separation $|\bar{y}^{\hat{o}}|-|\bar{y}^{\hat{i}}|$ increases with $r>r_J^{\hat{o}}$. But they approx to the axis in the region close to the instability  launch point $r_J$,  and for higher specific angular  momentum $|\ell|$. Though, in this model  of axisymmetric fluids with  constant specific angular  momentum, there is no ``collimation'' of the funnels or, if $x$=0 is the rotation axis, it is  $\partial_{|\bar{y}|}|\bar{x}|>0$ on the equatorial plane, where $|\bar{x}|>r_J$. Nevertheless any matter in open funnels,  with specific angular  momentum in $]\ell^{i},\ell^{o}[$,  will be bounded by the couple of configurations $O_x^{\hat{i}}\prec O_x^{\hat{o}}$. Then   $\partial_{x} (y_{\hat{o}}-y_{\hat{i}})>0$ (on  the section $x>0$ and $y>0$), the minimum value of the distance $(y_{\hat{o}}-y_{\hat{i}})$ is reached in the equatorial plane, where $x=0$.}}
\subsection{Restrictions on the existence of the open configurations: {limiting} surfaces}\label{Sec:det-open-in-ext}
 The set of  $O$ and $B_{ext}$ configurations which are   not related to  the critical points of the effective potential  correspond to the solutions  $\Pi=0$ of:
\be\label{Eq:Pi.I}
\Pi(\ell)={g_{\phi \phi}+2 \ell g_{\phi t} +\ell^2g_{tt}},
\ee
-- see Figs\il\ref{Figs:LS01POLOt},\ref{Figs:ThinkPi}.
%%
%\begin{figure}[h!]
%\begin{center}
%\begin{tabular}{c}
%\includegraphics[scale=0.3]{CblaPlot}
%%
 The effective potential,  related to the four-velocity component  $u_t=g_{\phi t} u^{\phi}+g_{tt} u^{t}$,
is  not well defined on the zeros of $\Pi(\ell)$.

Decomposing explicitly the function $\ell$ in terms of the  quantities   $\Sigma=u^t$ and  $\Phi= u^{\phi}$ then, the quantity  $\Pi$ can be written as
$\Pi(\Sigma, \Phi)=g_{tt} (\Sigma) ^2+2 g_{\phi t} \Sigma\Phi +g_{\phi\phi}\Phi^2$.
Thus,  $\Pi$ is related to the normalization factor  $\gamma$ for the stationary observers,  establishing thereby the light-surfaces (for example \cite{poisson,Pugliese:2018hju}).
Alternatively, by expressing the effective potential in terms of $L(\ell)$,  and by using definition Eq.\il(\ref{Eq:after}) and definition of  $\ell$  in Eq.\il(\ref{Eq:scond-d}),  one  obtains
$\Pi(L, E)= {E}^2 g_{\phi\phi}+2 {E} g_{\phi t} L(\ell)+g_{tt}  L(\ell)^2$.

The    attractor-ringed-disk system shows various  symmetry properties  with respect to the rationalization $\bar{\ell}=\ell/(a \sigma)$ and  $R=r/a$, where $\sigma=\sin \theta$ \cite{pugtot}. These quantities are dimensionless and, assuming $a>0$ with  $\ell$ positive or negative according to the fluids rotation, the parameter   $\bar{\ell}$, takes care of the symmetry for reflection on the equatorial plane through $\sigma$. Noticeably, many  properties of the \textbf{RAD} depend mainly on the rationalized specific angular momentum  $\bar{\ell}$. We discuss this kind  of symmetry in Sec.\il(\ref{Sec:div-spee}).  To characterize the  dependence on $a$, and the \textbf{RAD}  symmetry properties with respect to the equatorial plane, it  is convenient to rewrite the {quantity  $\Pi$ in terms of  $\bar{\ell}$ and $\bar{\bar \ell}\equiv \bar{\ell}/\sigma$ as follows}
\bea\nonumber
&&
\Pi(\bar{\bar{ \ell}},R)=-a^3 \sigma ^2 \left[2 (\bar{\bar \ell}-1)^2 R \sigma ^2+a \left(1+R^2-\sigma ^2\right) \left(1+R^2-\bar{\bar \ell}^2 \sigma ^2\right)\right],
\\&&\label{Eq:pipibarbar}
\Pi(\bar{\ell},R)=-a^3 \sigma ^2 \left[2  (\bar{\ell}-\sigma )^2R+a \left(1+R^2-\sigma ^2\right) \left(1+R^2-\bar\ell^2\right) \right],
\eea
{in dimensionless quantities.}
At fixed specific angular momentum $\ell$,  the zeros of the $\Pi$ function  define  \emph{limiting surfaces} of  the fluid configurations--Figs\il\ref{Figs:LS01POLOt},\ref{Figs:ThinkPi}.
For fluids with specific angular momentum $\ell \in \mathbf{L3}$, the limiting surfaces are  the cylinder-like  surfaces  $O_{ext}$, crossing the equatorial plane on a point which is  increasingly far from the attractor with increasing  specific angular momentum magnitude. A second $B$-like surface, embracing the \textbf{\textbf{BH}}, appears,  matching  the outer surface at the cusp $r_{\gamma}$.
Decreasing the specific angular momentum magnitude $\ell$ towards the liming value  $\ell_{\gamma}$,
the surface $ y(x)$, symmetric with respect to the equatorial plane, has a minimum at
$x=0$-- Eq.\il(\ref{Eq:pipibarbar}).
Then, for  $\ell=\ell_{\gamma}$ a cusp appears together  with a  inner surface closed on the \textbf{BH}.

The morphology of these surfaces   is  analogue to  the matter funnels of the open topologies having  $\ell\in\mathbf{L2}$ and $K\in\mathbf{K1}$. In this sense, the light-surfaces, for  $\ell=\ell_{\gamma}$, can be interpreted as ``limiting surfaces'' of the  open Boyer  surfaces.
Decreasing the specific angular  momentum $\ell$, from a starting $\ell\in \mathbf{L3}$, the $O_{ext}$ surfaces open up at the  radius $r_{\gamma}$ in $r<r_{\gamma}$, and  approaching the horizon as a $O_{in}$ topology towards the rotation axes as shown in Figs\il\ref{Figs:LS01POLOt},\ref{Figs:ThinkPi}. %Thus the  matter configurations,  at constant  specific angular  momentum, admit   the  unstable phases with cusps.
Comparing with the corotating surfaces, because of their rotation with respect to the black hole, the light-surfaces   for the counterrotating  fluids  form in  more distant  orbital regions. This fact implies the emergence of  a broad diversification of the unstable rotating structures for these fluids--see Fig.\il\ref{Figs:LS01POLOt}.

More generally, we can introduce the   {limiting geometric  surfaces}, or  \emph{$\gamma$-surfaces}, related to the {geometric properties of the Kerr spacetimes},    formed by  the surfaces  given by  the  solution of Eq.\il(\ref{Eq:scond-d}) and associated to the  specific  angular momentum $\left.\{\ell_{mso}^{\pm},\ell_{mbo}^{\pm},\ell_{\gamma}^{\pm}\}\right|_{\pi/2}$.

On the other hand, the   solutions of $\Pi(\ell)=0$, for  fixed  parameters   $\ell$ and $a/M$, define  the limiting hydrostatic surfaces, \emph{$h\gamma$-surfaces}, (with the notation $()^{\ell}$),  which are associated with  each hypersurface $\Sigma_{\ell}$,    whose topology and morphology changes
with the variation of one of the two rotational parameters $(\ell,a/M)$.
 More specifically, the $h\gamma$-surfaces are the limiting  surfaces to which the fluid configuration belonging to the same topological class, and determined by the  Euler problem on  $\Sigma_{\ell}\cup\Sigma_{a}$, approaches, at the variation of  the free $K$-parameter.
The $\gamma$-surfaces, on the other hand, constraint   the fluid  at variation of $\ell$.
At $\Sigma_{a/M}$, the $\gamma$-surfaces are the limits  of the $h\gamma$-surfaces, approached by varying $\ell$;  the $h\gamma$-surfaces  in turn limit the matter fluid surfaces as described  below. {In  Fig.\il\ref{Figs:LS01POLOt} an example of an $\ell$corotating sequence of  counterrotating  $h\gamma$-surfaces, the inner  one $O_{l}^{i_+}$ and outer $O_l^{o_+}$ one,  is shown.}

The \emph{$\gamma$-surfaces} define, for
the $\ell$counterrotating sequences on a $\Sigma_{a/M}$,
three geometric regions depending on  the attractor  spin:
\textbf{1.}
an {external} region, or $r> r_{\gamma}^+$, confined by the $\gamma$-surface $O_x^{\gamma_+}$ associated to the  counterrotating photon-orbit with a cusp at $r_{\gamma}^+$. \textbf{2.}
the region in  $]r_{\gamma}^-,r_{\gamma}^+[$,  and   finally \textbf{3.}  an
{internal}  region at $]r_h, r_{\gamma}^-[$. The open $\gamma$-surface $O_x^{\gamma_-}$ is cusped in $r_{\gamma}^-$,  and it corresponds to the  corotating photon orbit,
see Fig.\il\ref{Figs:LS01POLOt}.
The two critical surfaces $O_{x}^{\gamma_{\pm}}$, have one lobe closed on the black hole and the location on the corotating closed lobe of $O_{x}^{\gamma_-}$ is \emph{inside} the $O_x^{\gamma_+}$ configuration and, for $a<a_1=0.707107M$  this is inside the ergoregion of the Kerr spacetime, see \cite{ergon}.

Since there is  $\ell_{mbo}\in]\ell_{mso},\ell_{\gamma}[$, the $\gamma$-surfaces associated to the  the  specific angular momenta $\ell_{mso}$ and $\ell_{mbo}$, belong to the   internal region, and thus  they are configurations of the $O_{in}^{\gamma}$ type--see Figs\il\ref{Figs:Plotcredregrelre3D},\ref{Figs:Plotcredregrelre},\ref{Figs:LS01POLOt},\ref{Figs:ThinkPi}.  There are no   $\gamma$-surfaces in the external region.
For  the specific angular momentum $\ell_i$  on each hyperplane $\Sigma_{a/M}$, these surfaces
are  in turn determined a priori on each $\Sigma_{a/M}$,
whatever the specific angular momentum $\ell$ of the fluid is on that plane.
 The $h\gamma$-surfaces have a cusped topology  on the equatorial plane
at  $r_{\gamma}^{\pm}$  for counterrotating and corotating fluids respectively.
 We note that the investigation of these regions is  useful in particular  for  the  analysis  of the
  $\ell$counterrotating sequences at each $\Sigma_a$.
In fact, the limiting hydrostatic  surfaces  are never  cusped but at $\ell=\ell_{\gamma}$, where we have the topological class $O^{\ell}_x$, which coincide  with the
geometric light surfaces, or  $O_x^{\gamma}\equiv O_x^{\ell}$, closed on the black hole and opened outwards.
For the specific  angular momentum $\ell=\ell_{\gamma}^{\pm}$, the {matter} configurations, limited by  $O_x^{\gamma}\equiv O_x^{\ell}$,   can  be closed, centered in $r_c>r_{mso}$ for $K\in \mathbf{K0}$, or can be open in  $O_{ext}\succ O_{x}^{\gamma}$ for $K\in \mathbf{K1}$. The surface $O_{x}^{\gamma}$ is  a boundary surface reached by lowering the specific angular  momentum towards $\ell_{\gamma}$, or
\be\label{Eq:numer-box}
\lim_{|\ell|\rightarrow^+|\ell_{\gamma}|} O_{ext}\approx O_{x}^{\gamma}, \quad r_{ext}>r_{\gamma},\quad O_{ext}\succ O_{x}^{\gamma}, \quad |\ell_{ext}|>|\ell_{\gamma}|,
\ee
 where $r_{ext}$ is the
crossing point of $O$  on the equatorial plane\footnote{In these cases the sequentiality   will be intended according to the ordered sequences of  the equatorial crosses  $r_{ext}$ for the open, not-cusped, $O_{ext}$ surfaces.}. This is not a cusp but  it is a regular,  minimum   point
of the curve $y(x)$.
Therefore, the matter surfaces $O_{ext}$  have a critical point of the hydrostatic pressure, a maximum  in $r_{x}$ but, like  the correspondent $h\gamma$-surface  $O^{\ell}_{ext}$, it is open and  regular.

Concluding, there are three classes  of open matter configurations, $O_i\in\{O_x,O_{in},O_{ext}\}$, bounded  by the limiting hydrostatic  surfaces  $O_i^{\ell}\in\{O^{\ell}_x,O^{\ell}_{in},O^{\ell}_{ext}\}$ respectively, where $O^{\ell}_{x}\equiv O_{x}^{\gamma}$.
Each class  of limiting  surface $O^{\ell}_i$ is bounded by the correspondent open $h\gamma$-surface, $O_{i}^{\gamma}$,  and it approaches $O_{i}^{\gamma}$ varying  $K$.
Therefore
the two classes $O_{ext}^{\ell}$ and $O_{in}^{\ell}$ of open surfaces are separated by $O_x^{\ell}=O_{x}^{\gamma}$, so that there is
\bea\label{Eq:inclu-esclu}
&&O_{in}^{\ell_{in}}\prec O_{in}(\ell_{in})\prec O_x^{\ell}=O_{x}^{\gamma}\prec O_{ext}^{\ell_{ext}}\prec O_{ext}({\ell_{ext}}),
\\
&&\nonumber\mbox{with}\quad \ell_{in}<\ell_{\gamma}<\ell_{ext}\quad\mbox{and}
\\\label{Eq:zero-spotted}
&& O_x^{\ell}=O_{x}^{\gamma}\prec O_x(\bar{\ell})\prec O_{ext}^{\ell_{ext}}\prec O_{ext}({\ell_{ext}})
\\
&&\nonumber\mbox{with}\quad  \bar{\ell}<\ell_{in}<\ell_{\gamma}<\ell_{ext}.
\eea
On the other hand,
there are the  limiting specific  angular  momenta at
  $\ell_{mbo}>\ell_{mso}$, leading both to $O_{in}^{\gamma}$: or  $ O_{in}^{\ell}(\ell_{mbo})=O_{in}^{\gamma}(\ell_{mbo})\succ O_{in}^{\gamma}(\ell_{mso})=O_{in}^{\ell}(\ell_{mso})$.

The matter surfaces $O_x$  are bounded by the limiting $O_x^{\gamma}$, solution of $\Pi=0$.
The matter surfaces $O_{ext}\succ O_x^{\gamma}$
and  $O_{in}\prec O_x^{\gamma}$
are bounded by $O_x^{\gamma}$,  with specific angular momentum fixed in $\ell_{\gamma}$. They are also  bounded by the
 solutions of $\Pi(\ell)=0$, for the same  specific angular  momentum, and having  equal  topology
$O_{ext}\succ O_{ext}^{\ell}\succ O_x^{\gamma}$.
The $\gamma$-surfaces are approached by changing  the specific angular  momentum (see Eq.\il(\ref{Eq:inclu-esclu})). While the  $h\gamma$-surfaces are generally approached by an asymptotic limit of $K$, as  it is   clear from   Fig.\il\ref{Figs:Plotcredregrelre}.

We note that  the second and third  inequality  in Eq.\il(\ref{Eq:inclu-esclu}) (and the first and second of Eq.\il(\ref{Eq:zero-spotted}))
are ensured by  the relations among the  specific angular  momentum of the $\ell$corotating sequences,  as the following  general relations hold
\bea\label{Eq:tinEB}
&&
\partial_{\mathbf{Q}}r_{crit}^{[\mp]}\lessgtr 0,\quad r_{crit}^{[+]}=r_{min}=r_{cent},\\
&& r^{[-]}_{crit}\in\{r_{Max}=r_x, \; r_{Max}=r_{J},\; r_{int}, r_{ext}\},\quad \mathbf{Q}\in\{|\ell|, K\},
\eea
%.
 in particular for    $\mathbf{Q}=|\ell|$, where
Eq.\il(\ref{Eq:tinEB}), for $\mathbf{Q}=|\ell|$, does not apply to
the first and last inclusion relations of Eq.\il(\ref{Eq:inclu-esclu})
because the two couples of open  surfaces
$(O_{ext}, O_{ext}^{\ell})$ and $(O_{in}, O_{in}^{\ell})$, respectively, have the same topology and the same specific angular  momentum, the limit indeed is approached changing  the $K$ parameter or for $\mathbf{Q}=K$.
Thus,  the non-cusped  limiting  surfaces  can be
the  regular couples  $B^{\ell}_{ext}<O^{\ell}_{ext}$,
when $\ell\in \mathbf{L3}$, that is  $\ell>\ell_{\gamma}$, in the external region, according to Figs\il\ref{Figs:Plotcredregrelre},\ref{Figs:LS01POLOt},\ref{Figs:ThinkPi}.

Associated with these configurations, there is   the couple   $B_{ext}<O_{ext}$ in $\mathbf{L3\cup K1}$.
Otherwise  there can be, in the internal region, also a $O^{\gamma}_{in}$  surface for specific angular momentum $\ell<\ell_{\gamma}$ embracing the horizon. The $O_{in}$ surfaces are due to  the cusp opening  occurring when the  magnitude of the specific angular  momentum decreases  (where  $\partial_y^2 x>0$).
 There are therefore the $h\gamma$-surfaces $O^{\ell}_{in}$, approaching the proper limits on the specific angular  momentum (starting by initial data in $\mathbf{L0}$, $\mathbf{L1}$ or $\mathbf{L2}$) the $\gamma$-surfaces    $O^{\gamma}_{in}$,  zeros of  Eq.\il(\ref{Eq:Pi.I}) for $\ell=\{\ell_{mso},\ell_{mbo}\}$.
Indeed, \emph{decreasing} $\ell_0\in \mathbf{L3}$, we have  the $h\gamma$-surfaces sequences:
$ \{\left.(B^{\gamma}_{ext}<O^{\gamma}_{ext})\right|_{\mathbf{L3}}, \left.O^
{\gamma}_x\right|_{\ell_{\gamma}}, \left.O^{\gamma}_{in}\right|_{\ell<\ell_{\gamma}})\}
$.
For the  open crossed surfaces  $O_x$ (for $\ell\in \mathbf{L2}$) there are
\be
\lim_{\ell\rightarrow^+\ell_{\gamma}}O_x\approx O^{\gamma}_x=O^{\ell}_x,\quad O_x\succ O^{\gamma}_x\quad r_
{x}=r_{J}>r_{x}^{\gamma},\quad  O_x< O_x^{\gamma},
\ee
see  also Eq.\il(\ref{Eq:numer-box})  and Fig.\il\ref{Figs:LS01POLOt}.
The open surfaces
$O_{ext}$ are limited by the configurations  $O^{\gamma}_{ext}$, in  other words $r_{ext}=y_{3}>r_{ext}^{\gamma}$, where
 $O_{ext}> O_{ext}^{\gamma}$--Fig.\il\ref{Figs:LS01POLOt}.
There  is also $B^{\gamma}_{ext}<O^{\gamma}_{ext}$,
and
$
\partial_{|\ell\in \mathbf{L3}|}B^{\gamma}_{ext}>0$, $
\partial_{|\ell\in \mathbf{L3}|}r^{\gamma}_{ext}>0
$.
For  specific angular momentum $\ell<\ell_{\gamma}$,   there are the  open surfaces $(O^{\gamma}_{in},O^{\ell}_{in},O_{in})$.%,
 Whereas, for
 $\ell \in \mathbf{L2}$,   there are the  surfaces $\{O_{x},O_{in}, B_{in}\}$. The   $B_{in}$ configuration occurs occurs for  $K<K_{min}$ (then also $K\in \mathbf{K_*}$) or  $K\in]K_{Max},1[$. At fixed  $\ell\in \mathbf{L1}$, there can be the inner $B_{in}$ in $\mathbf{K_*}$: i.e.,  increasing  $K>0$,   there is the  sequence $\{B_{in},(B_{in},C),C_x,O_{in})\}$. If the starting point is  $\ell\in \mathbf{L0}$, then, increasing $K>0$, there is the  sequence
  $\mathcal{B}_{K}=\{B_{in},O_{in}\}$. In other words, the specific angular  momentum is low enough to not to lead to the formation of an outer lobe, but it eventually opens in $O_{in}$.
The  $B_{ext}$,  associated to $O_{ext}$ for high values of the specific angular momenta,  has    morphology similar to the $B_{in}$ surface.
  The $B$ configurations for  $\ell\in\mathbf{L2}$, where the cusped surfaces $O_x$ can appear, are classified as  $B_{in}$, in fact the $B_{ext}$ ones  are separated by $O_{ext}$ for each  value of $\ell \in \mathbf{L3}$ and any  $K$ value--see Fig.\il\ref{Figs:ThinkPi} and Fig.\il\ref{Figs:Plotcredregrelre}.

For parameter  $K\in \mathbf{K0}>K_{Max}$, there are $B_{in}$ surfaces opening   for
 $K\geq1$ as $O_{in}\in] O^{\gamma}_{in},O^{\gamma}_x[$ and,
similarly to $O_{in}$ with $\ell\in \mathbf{L1}$,
and  $O_{in}$ with $\ell \in \mathbf{L0}$.
The presence of a minimum point of the hydrostatic pressure always implies, in a Kerr black hole geometry,  the presence also of a maximum pressure point. The inverse is not true, for example for  $\ell\in\mathbf{L3}$,
when there is only one  family of non-critical open surfaces $O_{ext}$ at  $r>r_{\gamma}$.

We summarizing saying that  for $K\in \mathbf{K1}$ there are open surfaces  for any  specific fluid  angular momentum $\ell$.
The cusped open configurations  $O_x$,  closed on the  \textbf{BH},  are associated to parameters $\ell\in \mathbf{L2}\cup K\in \mathbf{K1}$, where the limiting surface is
\bea&&
{K}\in \textbf{K1}\quad O_x\succ O^{\gamma}_x \quad \quad \ell\in \mathbf{L2}  \quad r_J\in ]r_{x}^{\gamma}=r_{\gamma},r_{mbo}]\\
&& \partial_{|\ell|} r_{J}<0\quad \partial_{K} r_{J}<0,\quad\lim_{\ell\rightarrow^+ \ell_{\gamma}}O_x\approx O^{\gamma}_x.
\eea
 However, we have
\be\lim_{K\rightarrow\infty}O_{ext}\approx O_{ext}^{\ell}\quad\mbox{ while }\quad\lim_{\ell\rightarrow^+\ell_{\gamma}}O_{ext}^{\ell} \approx O^{\gamma}_x.
\ee
The open surfaces, of    $O_{in}$  class  with  $\ell\in\mathbf{L2}$ are limited by $O_{in}^{\ell}$ at equal  $\ell$. These are limited by the boundary surfaces with specific angular momentum
$\ell_{mbo}$ and  $\ell_{mso}$.
 However, as clear from the Fig.\il\ref{Figs:LS01POLOt},
for  $x$ large enough, the funnels of $h\gamma$-surfaces go far from the source and, independently by the magnitude $\ell_{a}/\ell_b=-1$, the  two $\ell$counterotating funnels are getting closer.%\rtb{AXES}.

%%%%%%%%%%%%%%%%%%%%%%%%%%%%%%%%%%%%%%%%%%%%%%%%%%%
On the other hand,  for sufficiently high  magnitude  of the specific angular momenta, i.e.      $\ell in \mathbf{L3}$ and $K\in\cup K0$, there are  maximum pressure points but not  the  minimum of the hydrostatic pressure.  Consequently there are  closed stable $C$ or open  $O_{ext}$ topologies.
The Paczy\'nski-Wiita instability can occur only after a reduction of the specific angular  momentum, when the disk  center approaches the attractor.
 If the  specific angular  momentum decreases  from  the  $\ell\in\mathbf{L3}\cup K\in K0$ to $\ell\in \mathbf{L1}\cup K\in{K0}$,  then  the unstable phase  will necessarily correspond  to an accretion. If the decrease of specific angular  momentum occurs with an increase of $ K $,  then it can give rise to the  open cusped   $O_x$  surface.
The high values of the specific angular momentum, $\ell\in \mathbf{L3}$ or also  $\ell\in \mathbf{L2}$, are associated to closed equilibrium configurations or open surfaces.  If the disk stretches sufficiently, the pressure at the inner edge  can reach the proper minimum  value to be open. On the other hand if   there is $\ell\in\mathbf{L3}$, then  the  elongation of the configurations and and fluid angular  momentum magnitude are too high causing  the  opening of a  $O_{ext}$ surface.
\section{Concluding Remarks}\label{Sec:Conclusion}
We focused on  the open, unstable solutions of Euler equations for  \textbf{RADs},  aggregations of  several tori of one-species particle  perfect fluid centered  on the equatorial plane of  a Kerr \textbf{SMBH}.
These  structures, have been variously associated with jet of proto-jet emission.
This investigation  then fits into the more broad discussion on the role and significance of open surfaces in relation to  (matter) jets emission and collimation, as well as jet-accretion correlation--see also \cite{KJA78,AJS78,Sadowski:2015jaa,Lasota:2015bii,Lyutikov(2009),Madau(1988),Sikora(1981)}.

{The issue of the   location of the inner edge of a single torus is a  relevant aspect of the  possible jet-accretion correlation, relating jet-emission to the inner part of the accretion disk--see for a discussion of the inner edge problem
    \cite{Krolik:2002ae,BMP98,2010A&A...521A..15A,Agol:1999dn,Paczynski:2000tz}. }
    {Attempts to characterized  narrow, relatively long  matter funnels of jets, here considered as proto-jets configurations, are for example in
 \cite{NatureMa,Maraschi:2002pp,Chen:2015cga,Yu:2015lqj,Zhang:2015eka,Sbarrato:2014uxa,MSBNNW2009,Ghisellini:2014pwa,BMP98}. For updated investigations  on jet emission, detection and jet-accretion disk  correlation see for example
\cite{liska,Caproni:2017nsh,Inoue:2017bgt,Gandhi:2017dix,Duran:2016wdi,Vedantham:2017kyb,Bogdan:2017pgi,Banados:2017unc,DAmmando:2017ufp}
and  also
\cite{NL2009,Fender:2004aw,Fender:2009re,Soleri:2010fz,Tetarenko18,2017ApJ...851...98T,FangMurase2018}.
}

After  introducing  the  model  for the single    torus orbiting  of the  \textbf{RAD}, in Sec.\il (\ref{Sec:models}) we discussed the main features of the ringed accretion disks, and  the emergence of the  different  instabilities for the systems.
The problem to find the  solutions of the Euler equations for the \textbf{RADs} has been reduced to the study of the critical points for  an  effective potential describing  the orbiting fluids with proper boundary conditions; in this way we could  estimate very precisely the diverse contributions of the background and the centrifugal forces on dynamics of the  single torus as well as on the set of configurations in the \textbf{RAD} and the entire agglomeration.
In  Sec.\il(\ref{Sec:div-spee}), we have deepened this aspect by taking advantage of system symmetries, considering a different parametrization for the effective potential function highlighting the symmetry of motion and background represented by Kerr axi-symmetric solution. Eventually we introduced different rotational parameters.
This has allowed us to highlight the role of the dimensionless  radius $R\equiv r/a$ and of the fluid specific angular momentum with respect to the black hole spin,  the  ratio $\ell/a \sin \theta$,  pointed out also  in \cite{pugtot}, and  the quantities  $\mathcal{A}^{\pm}=\ell\pm a$.

The analysis has ultimately singled out  the role of the  limiting surfaces, the  \emph{$\gamma$-surfaces} and \emph{h$\gamma$-surfaces}, for the open configurations. We proved there is a strict correlation  between   different $\gamma$-surfaces (geometric surfaces of the spacetime structure)
and  the  $h\gamma$-surfaces having roles in the matter models. The $h\gamma$-surfaces limit the  fluid configuration, while
the $\gamma$-surfaces constraint   the fluid,  at variation of $\ell$.
The   {limiting geometric  surfaces},  \emph{$\gamma$-surfaces}, are  related to the {geometric properties of the Kerr spacetimes} inherited  by  the  solution of Eq.\il(\ref{Eq:scond-d}), associated to the  specific  angular momentum $\left.\{\ell_{mso}^{\pm},\ell_{mbo}^{\pm},\ell_{\gamma}^{\pm}\}\right|_{\pi/2}$.
Given the significance  from the phenomenological point of view of different   instabilities  which characterize the toroidal population in the  \textbf{RAD}, it  is clear that a study of these constraints (especially in relation to the central attractor) is important.
 The limiting hydrostatic surfaces, \emph{$h\gamma$-surfaces}, are associated with  each hypersurface $\Sigma_{\ell}$,   of constant specific fluid angular momentum,  whose topology and morphology changes
with the variation of one of the two  system rotational parameters $(\ell,a/M)$.
At $\Sigma_{a/M}$, the $\gamma$-surfaces are the limits  of the $h\gamma$-surfaces, approached by varying $\ell$;  the $h\gamma$-surfaces  in turn limit the matter fluid surfaces.

The role of these surfaces in the \textbf{RADs}, their origins and the destabilizing effects on the system were briefly addressed,  possible magnetic effects in a magnetized \textbf{RAD} contests are discussed in \cite{Pugliese:2018zlx},
while
a more thorough discussion of the boundary  conditions
for the  open solutions   are postponed to  future analysis.
 We expect these considerations can play an important part in the  phenomenological aspects connected  with  X-ray emission in \textbf{SMBH-AGNs} and in the study of the  possibility for a  jet-accretion correlation.
\begin{acknowledgements}
D. P. acknowledges support from the Junior GACR grant of the Czech Science Foundation No:16-03564Y.
Z. S. acknowledges  the Albert Einstein Centre for Gravitation and Astrophysics supported by grant No.
 14-37086G.
 \end{acknowledgements}

\end{document}